\documentclass[11pt,nofootinbib,preprint,tightenlines,amsmath,amssymb]{revtex4}

\newcommand{\beq}{\begin{equation}}
\newcommand{\eeq}{\end{equation}}
\newcommand{\be}{\begin{equation}}
\newcommand{\ee}{\end{equation}}
\newcommand{\beqa}{\begin{eqnarray}}
\newcommand{\eeqa}{\end{eqnarray}}

\newcommand{\bean}{\begin{eqnarray*}}
\newcommand{\eean}{\end{eqnarray*}}
\newcommand{\RR}{\mathbb{R}}
\newcommand{\ZZ}{\mathbb{Z}}
\newcommand{\extd}{{\mathrm {d}}}

\newcommand{\pa}{\partial}

\newcommand{\OO}{\mathcal{O}_{\Gamma}}
\newcommand{\VV}{\mathcal{V}}
\newcommand{\mc}[1]{{\mathcal{{#1}}}}

\newcommand{\C}[4][x]{C_{{#1}_{#2} {#1}_{#3} {#1}_{#4}}^{\rho_{#2}\rho_{#3}\rho_{#4}}}

\usepackage{psfrag}
\usepackage[dvips]{graphicx}
\newtheorem{lemma}{Proposition}

\begin{document}

\title{
Hidden Quantum Gravity in 3d Feynman diagrams}

\author{Aristide Baratin$^1$\thanks{email: abaratin@ens-lyon.fr},
Laurent Freidel$^{1,2}$\thanks{email:
lfreidel@perimeterinstitute.ca}}

\affiliation{\centerline{\footnotesize \it
$^1$Laboratoire de Physique, \'Ecole Normale Sup{\'e}rieure de Lyon} \\
\centerline{\footnotesize \it 46 all{\'e}e d'Italie, 69364 Lyon
Cedex 07, France.}\\
\centerline{\footnotesize \it $^2$Perimeter Institute for
Theoretical Physics} \\
 \centerline{\footnotesize \it 31 Caroline street North, N2L2Y5
Waterloo, ON, Canada}}

\begin{abstract}

In this work we show that  3d Feynman amplitudes of standard QFT in flat and  homogeneous space
can be naturally expressed as expectation values of a specific  topological spin foam model.
The main interest of the paper is to set up a framework which gives a background independent perspective on
usual field theories and can also be applied in higher dimensions.
We also show that this Feynman graph spin foam model, which encodes the geometry of flat space-time, can be
purely expressed  in terms of algebraic data associated with the Poincar\'e group.
 This spin foam model turns out to be the spin foam quantization of a BF theory based
on the Poincar\'e group, and as such is related to  a quantization of 3d gravity in the limit $G_{N}\to 0$.
We investigate the 4d case in a companion paper \cite{4d}  where the strategy proposed here  leads to similar results.

\end{abstract} \maketitle

\tableofcontents

\newpage

\section{Introduction}
There has been a recent understanding and  new results concerning the consistent coupling of matter fields to 3d quantum gravity amplitudes, in the context
of the spin foam approach to 3d quantum gravity \cite{PR1,PR2,B1,B2,B3,PN}.
These results have led to the construction of an effective field theory, describing the coupling of quantum fields to quantum gravity
\cite{PR3,EFT}, which arises after the exact integration of quantum gravitational degrees of freedom.
This effective field theory lives on a non-commutative space-time whose geometry is studied in detail in \cite{MF}.
When the no gravity limit $G_{N}\to 0 $ is taken, the geometry becomes commutative and usual field theory is recovered.

From the spin foam side, the way usual Feynman diagrams are recovered is highly non trivial, since quantum gravity amplitudes are
expressed as purely algebraic objects in terms of spin foam models, a language quite remote from usual field theory language.
The fact that Feynman diagrams arise from this picture effectively, amounts to show that usual field theory can be expressed in a
background independent manner, where flat space-time emerges dynamically from the choice of spin foam amplitudes.
Namely, this amounts to show that Feynman amplitudes can be expressed as the expectation value of certain observables in a topological spin foam model.

Note that this last statement is not a statement regarding quantum gravity but one regarding well known Feynman diagrams in 3d.
It is then natural to wonder if one can reach a deeper understanding of this property; whether it is tied up to the topological nature
of gravity in dimension 3 or if it can be extended to field theory in higher dimensions. It also begs the question, whether or not
Feynman diagrams contain some seeds of information about quantum gravity dynamics.

    A natural strategy to address successfully these questions is to start from a careful study of the structure of Feynman
graph amplitudes, without any assumption about the nature of quantum gravity dynamics. The goal is to see why and how such familiar
objects can be interpreted in terms of spin foam models, which have appeared to be a suitable language to address dynamical question
in a background independent approach to quantum gravity \cite{rev}.

Usual Feynman diagrams are natural observables\footnote{Of course, any observable quantity is actually given by a sum of Feynman diagrams, for instance those of fixed valency and given degree, as generated by a field theory at given order of
perturbation, and subject to renormalization procedure. In order to keep the exposition simple we
will refer to individual Feynman diagrams, in a regularized form in order to avoid divergences.}
from which physical predictions regarding matter interactions in
space-time can be extracted. As such, they allow one in principle
to probe the geometry of space-time at arbitrarily small scales.
Moreover, they can be naturally promoted to physical observables in
any theory of quantum gravity. Therefore, the preferred spin foam
model - if any - which is related to Feynman diagrams should give us
some important seeds of information about the structure of quantum
gravity amplitudes. In other words, if one assumes that it is
possible to consistently describe quantum gravity amplitudes coupled
to matter in terms of a spin foam model, a necessary requirement for
the spin foam amplitudes, in order to be physically relevant, is to
reduce to the ones appearing in the study of Feynman diagrams when
$G_{N}\to 0$. This gives strong restrictions on the admissible,
physically viable spin foam models.

In this paper we focus our study to the case of $3$-dimensional Feynman diagrams, but as should be clear from our work and
as shown in a companion paper \cite{4d},
most of the results obtained here can be extended to the more interesting and challenging case of $4$-dimensional amplitudes.
What makes the analysis work is not that  gravity is a topological theory but the fact that, when coupled to matter,
quantum gravity {\it in the limit } $G_{N}\to 0$ behaves as a topological field theory. This is a trivial fact in dimension $3$ and
has been argued to be true in dimension $4$ when one takes this limit while preserving diffeomorphism invariance \cite{AF}.

The main idea behind our analysis is the fact that
in 3d, quantum gravity amplitudes provide an integration measure for the integration of Feynman graph amplitudes.
This idea is contained in \cite{PR3} and spelled out in \cite{B4}.
When gravity is turned off, this measure should be related to the Lebesgue measure expressed in terms of relatives lengths,
which are Poincar\'e scalars.
When gravity is turned on, this measure is deformed and reveals the quantum nature of the
corresponding space-time.

The organization of this paper is as follows. In  section \ref{inv} we first express explicitly the change of variables from the
Lebesgue measure that arises in Feynman graph integration,
to the measure expressed in terms of relative distances.
In section \ref{result} we present our main result, namely that Feynman integrals can be expressed as the expectation value of an observable
for a topological spin foam model.
This model is constructed out by removing  all explicit information about flat space into a uniquely defined choice of amplitudes.
We carefully study this model, especially its symmetries and its gauge fixing properties, and show that it defines a topological state sum model.
This topological model, which naturally emerges from Feynman graphs, has already been studied in the mathematical literature from a different perspective
 by Korepanov \cite{Kor}.
In  section \ref{alg} we show that the tetrahedral weights entering the definition of the topological state sum model
can be given an algebraic interpretation in terms of the $6j$-symbol of the Poincar\'e group; this $6j$-symbol is computed explicitly
and for the first time.
We also relate this model to the Ponzano-Regge model via a doubling and limiting procedure.
Finally we show how to extend these results to the case of a non vanishing cosmological constant.

\section{Feynman diagrams in terms of invariant measure}\label{inv}

We consider QFT in flat Euclidean D-dimensional space-time and focus on closed Feynman diagrams for a scalar field.
The Feynman amplitude for a graph $\Gamma$ is given by an integral over the position in $\RR^D$ of N vertices. The (Euclidean) Poincar\'e
group $ISO_D$ acts diagonally on these $N$ points
\[ (\vec{x}_1,\cdots,\vec{x}_N)\to (\Lambda\cdot\vec{x}_1
+\vec{a},\cdots,\Lambda\cdot\vec{x}_N +\vec{a}) \]
where $(\Lambda, \vec{a}) \in SO_D \times \RR^D$ is a general  element of the Euclidean Poincar\'e group. The integrand
is a function of the distances between the vertices, namely an invariant under the action of $ISO_D$
\beq \label{intx}
I_{\Gamma} = \int_{\RR^D} \mathrm{d}^Dx_1\cdots\mathrm{d}^Dx_N\, \OO(|\vec{x}_i - \vec{x}_j|), \quad \mbox{with} \quad
\OO = \prod_{(ij)\in \Gamma}G^F(\vec{x}_i - \vec{x}_j)
\eeq
$G_F$ is the Feynman propagator and the product is over all edges of $\Gamma$.
Following the ideas of \cite{PR3,B4}, we want to express this integral in terms of
the invariant measure, involving the quantities $l_{ij}=|\vec{x}_i-\vec{x}_j|$. On one hand, $ND$ variables are integrated over in (\ref{intx})
while the dimension of the Poincar\'e group is $\frac{1}{2}\,D(D+1)$; consequently,  the invariant measure  depends on
$ND-\frac{1}{2}\,D(D+1)$ variables: the number of independent variables required to place the vertices with respect to each
other\footnote{Precisely this quantity is the number of \textit{continuous} variables required to place the vertices with
respect to each other modulo \textit{discrete} transformations.}. On the other hand, the number of length variables
$l_{ij}$ is $\frac{1}{2}N(N-1)$. Note that the quantity:
\beq
\label{nd}
\frac{N(N-1)}{2} - \left[ND-\frac{D(D+1)}{2}\right]= \frac{(N-D)(N-D-1)}{2}
\eeq
which computes the number of redundant edge lengths, vanishes only for $N=D$ and $N=D+1$.

First, for $N=D$, we split the measure into the Haar measure for $ISO_D$ and a product of $l_{ij}\extd l_{ij}$
\be \label{measD}
\extd^Dx_1\cdots \extd^Dx_D = \extd^D a \, \extd{\Lambda}\,\prod_{i<j}l_{ij} \extd l_{ij},
\ee
where $a$ is the barycenter of the $D$ points and $d\Lambda$ is the Haar measure on $SO_{D}$. The volume of $SO_{D}$ with respect to this measure
is given by
\[
A_{D}=\frac{2^{D-1}
\pi^{\frac{D(D+1)}{4}}}{\prod_{k=1}^D\Gamma(\frac{k}{2})},
\]
so $A_{2}= 2\pi, A_{3}=2^3\pi^2$ and $A_{4}=2^4\pi^4$. Given an additional point ${x}_{{D+1}}$, one can build a $D$-simplex in $\RR^D$ with
vertices ${x}_{i}$, and express the Lebesgue measure in terms of the edge lengths as
\be \label{measx}
\extd^Dx_{D+1} = \sum_{\epsilon=\pm 1}\frac{\prod_{i=1}^D l_{iD+1}\extd l_{iD+1}}{\mathcal{V}(l_{ij})}
\ee
$\mathcal{V}(l_{ij})$ is D! times the volume of the simplex whose edge lengths are $l_{ij}$.
In the rest of the paper, we will simply call `volume' the quantity $\mathcal{V}$. $\epsilon$ labels
the orientation\footnote{The notion of orientation is precisely defined in appendix \ref{geo}.} of the $D$-simplex $x_{1}\cdots x_{D+1}$.

The proof of (\ref{measx}) is easily obtained by computing the jacobian of the linear function which maps an orthonormal basis $e_{i}$ to the
basis defined by the vectors $\{x_{i}\}_{i=1\cdots D}$. Let $y \in \RR^D$ and $l_{i}=|y-x_{i}|$; we have
\beqa
\prod_{i=1}^D l_{i}\extd l_{i} &=& \prod_{i=1}^D \extd \frac{(y-x_{i})^2}{2}
=  \prod_{i=1}^D \extd y_{j} \mathrm{det}\left(\frac{\partial}{\partial y^j}\frac{(y-x_{i})^2}{2}\right) \\
&=&\prod_{i=1}^D \extd y_{i} \, |{\rm det}(e_{j}\cdot(y-x_{i}))| = \extd^{D}y  \sqrt{{\rm det}((y-x_{j})\cdot(y-x_{i}))}
\eeqa
This last determinant is $D!$ times the volume of the simplex $x_{1},\cdots,x_{D},y$. The previous identity is a local identity
showing that the map $y\to l_{i}$ is locally invertible away from the hyperplane spanned by $x_{1},\cdots, x_{D}$. This map is globally invertible on each
half space separated by this hyperplane. The sum over orientations comes from the fact that the
set of edge lengths $l_{i}$ determines the position of $y$ up to   a reflection with respect to the hyperplane spanned by $x_{1},\cdots, x_{D}$. Such a
reflection is in $O(D)$, but not in $SO(D)$, and changes the orientation of the $D$-simplex. The equality of measures
(\ref{measx}) is then understood as an equality for integrating out a function of $x_{D+1}$ in the LHS and a  function of $l_{iD+1}$
and orientation $\epsilon$ in the RHS. The proof of (\ref{measD}) is obtained by a direct computation and by recurrence.

So putting (\ref{measD},\ref{measx}) together we get, for $N=D+1$ points in $\RR^D$, that the invariant measure is given by:
\beq \label{invTet}
\mathrm{d}^Dx_1\cdots\extd^Dx_{D+1}\,=\extd^D a \, \extd{\Lambda}\,\sum_{\epsilon} \frac{\prod_{ij}l_{ij}\extd l_{ij}}{\mathcal{V}(l_{ij})}.
\eeq
A similar identity already appeared in \cite{LDA1}.

Now for $N=D+2$, the quantity (\ref{nd}) is equal to 1. Accordingly, the invariant measure will be formulated using all the
distances, except for one. One can indeed express it in terms of the edge lengths of the polyhedron formed by two D-simplices glued
together in $\RR^D$ along a $(D-1)$-simplex (as shown in Fig \ref{pent} for $D=3$).
The missing edge length is the one joining the only two vertices, denoted by $a,b$, which do not
belong to the common $(D-1)$-face. The measure is obtained by using twice (\ref{measx})
\be \label{measd2}
\mathrm{d}^Dx_1\cdots\extd^Dx_{D+2}=\extd^D a \,\extd{\Lambda}\sum_{\epsilon_{a},\epsilon_{b}} \frac{ \prod_{(ij)\neq (ab)}l_{ij}\extd l_{ij}}
{\mathcal{V}_{a}\mathcal{V}_{b}}
\ee
where the product is over all edges $(ij)$ different from $(a,b)$,  $\mathcal{V}_{a}$ denotes the volume of the $D$-simplex not containing $a$, and the
sum is over orientations of the $D$-simplices  $a,b$.

Note that the edge lengths $l_{ij},(ij)\neq(ab)$ only determine the
geometry of the polyhedron modulo reflections of the D-simplices along the common $(D-1)$-face.
In order to fully determine the geometry, one needs
to specify the orientations of the two $D$-simplices denoted $a$ and $b$.
Once these orientations $\epsilon_{a}, \epsilon_{b}$ are specified, the missing edge length can be uniquely reconstructed: $l_{ab}=l_{ab}(l_{ij},\epsilon)$.
For fixed $l_{ij}$, the missing distance $l_{ab}$ can take two values denoted by $l_{ab}^\pm$, with $l_{ab}^- \leq l_{ab}^+$, depending on whether
the two points $(a,b)$ are separated or not by the hyperplane spanned by the common face. A simultaneous change of the two orientations
is equivalent to a symmetry with respect to this hyperplane, under which all distances remain invariant. Therefore,
the length $l_{ab}$ depends only on the \textit{product}
$\epsilon_a\epsilon_b$, in such a way, according to our conventions, that $l_{ab}^\pm = l_{ab}^{\epsilon_a\epsilon_b}$.
The Fig \ref{convD2} illustrates the $D=2$ case. More details are given in appendix.

\begin{figure}[h]
\begin{center}
\includegraphics[width=10cm, height=4cm]{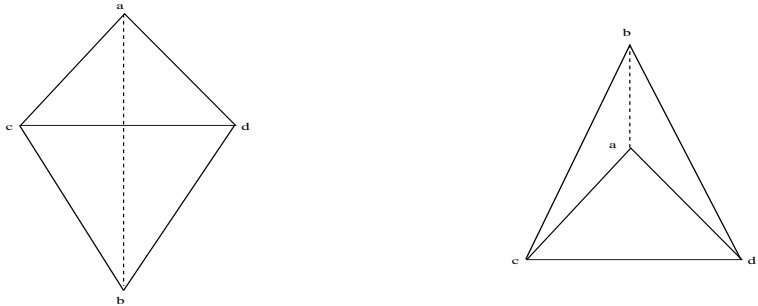}
\end{center}
\caption{Two triangles $acd$ and $bcd$ in the Euclidean plane share an edge $(cd)$. On the left (resp. right) $l_{ab} =  l_{ab}^+$  (resp. $l_{ab}^-$).}
\label{convD2}
\end{figure}

For a general value of $N=D+k$, we use repeatedly ($k$ times) the equation (\ref{measx}) in order to get the invariant
measure. This invariant measure is based on a triangulation which is constructed recursively:
  each time a vertex is added, we choose a $(D-1)$-face on the existing triangulation, and build a new $D$-simplex
  from this data. At each step we use (\ref{measx}) to construct the invariant measure whose form depends on the triangulation $\Delta_{k}$.
The triangulations obtained this way are of special type: they are triangulations with a boundary being topologically a $(D-1)$-sphere and such that
every $(D-2)$-face lies on the boundary.

  More formally, let $B_{D}$ be a $D$-dimensional ball $B_{D}$ with $n=D+k$ points marked on its boundary. We now consider a triangulation
  $\Delta_{k}$ of $B_{D}$ with $D+k$ vertices, such that every vertex and every $(D-2)$-face are on the boundary.
  In other words  $\Delta_{k}$ is a triangulation of $B_{D}$ which possesses $D+k$ boundary vertices and which does not
possess any internal $(D-2)$-face.
  If one draws the dual $2$-skeleton of $\Delta_{k}$, in which vertices are dual to $D$-simplices, edges to $(D-1)$-simplices and faces to
  $(D-2)$-simplices,
  one sees that our condition on $\Delta_{k}$ implies that this dual graph is a $D$-tree (a $D+1$ valent graph containing
no closed loop) with open ends.
  It can be checked that the number of edges of such a triangulation matches the number of Poincar\'e invariant quantities.
  Given such a triangulation the invariant measure reads
  \be
   \mathrm{d}^Dx_1\cdots\mathrm{d}^Dx_{D+k} = \extd^{D}a \extd\Lambda \sum_{\epsilon \in \{\pm 1\}^k} \prod_{e\in \Delta_{k}}l_e\extd l_e
   \prod_{\sigma\in\Delta_{k}}\frac{1}{\mathcal{V}_{\sigma}}
\ee
where the first product is over all edges $e$ of  the triangulation $\Delta_{k}$, the second product is over all its
$D$-simplices $\sigma$ and the summation is over all orientations of its $D$-simplices.

By construction the data $\{l_{e},\epsilon_{\sigma}\}$ defines uniquely a flat geometry on $\Delta_{k}$ and allow us to
reconstruct
the `missing' edge lengths, namely the length of the edges that do not belong to $\Delta_{k}$.
The set of distances $\{l_{ij}\}$
in the integrand of (\ref{intx}) splits up into the lengths $\{l_e,e\in\Delta_{k}\}$ and the `missing' distances denoted by $\{l^{\epsilon}_{e'}\}$.
These missing distances are $\epsilon$-dependent functions of the $l_e$.

The Feynman amplitude (\ref{intx}) is finally given by:
\beq
\label{Feyamp}
I_{\Gamma} = \int \prod_{e\in\Delta_{k}}l_e\extd l_e\,\sum_{\epsilon \in{\{\pm 1\}}^{k}}\,
\prod_{\sigma\in\Delta_{k}}\,\frac{1}{\mathcal{V}_{\sigma}}\, \OO(l_e,l_{e'}^\epsilon({l_e}))
\eeq
where we have dropped out the overall factor which corresponds to the gauge volume.

Let us conclude this part: we have started with a Feynman amplitude on flat space-time; in the usual
formulation, information about flat geometry is encoded in the Lebesgue measure. By working with the invariant measure, we
have moved  this information in the weight assigned to the triangulation, and in the relations $l_{e'}^\epsilon({l_e})$
between `missing' lengths, labels and orientations.
Our goal now is to show that this information about flat geometry can be encoded entirely into a dynamical topological state sum (spin foam) model.

\section{Feynman diagrams as spin foam models}
\label{result}
\subsection{The main result}

One of the main results of this paper is the fact that
the Feynman amplitude (\ref{Feyamp}) can be written as the expectation value of an observable
\beq \label{exp}
I_{\Gamma}=\langle\OO (l_e)\rangle_{Z_{\Delta}}, \qquad \OO = \prod_{e \in \Gamma} G^F(l_e)
\eeq
for the spin foam model:
\beq \label{3dmod}
Z_{\Delta}=\frac{1}{(2\pi)^{|e|}}\int\prod_{e\in\Delta}\mathrm{d}l_el^2_e\sum_{\{s_e,\epsilon_{\tau}\}}\left(\prod_{\tau}\,
\frac{e^{\imath \epsilon_{\tau} S_{\tau}(s_{e},l_{e})}}{\mathcal{V}_{\tau}}\right)
\eeq
$\Delta$ is an arbitrary  triangulation of a closed 3D-manifold  and $\Gamma$ is embedded into the one-skeleton of $\Delta$.
$|e|$ is the number of edges  of $\Delta$.
The edges carry a double label ($l_e, s_e$), where $l_e$ is a real positive number and $s_e$ is an integer.
These labels are summed over a domain where triangular inequalities are satisfied for the $l_{e}$'s. $\epsilon_{\tau}=\pm 1$ denotes the orientation
of the tetrahedron $\tau$. The product in parenthesis is over all tetrahedra. The action for each tetrahedron is similar to the Regge action \cite{Regge}
\be \label{action}
S_{\tau}(s_{e},l_{e})= \sum_{e\in \tau} s_{e} \theta_{e}^\tau(l_e),
\ee
with $\theta_{e}^\tau(l_{e})$ the interior dihedral angle of the edge $e$ in $\tau$. The observable is
simply the product of propagators in which the distances are replaced by the labels $l_e$ living on the graph $\Gamma$.
The equality (\ref{exp}) is obtained if one restricts to trivial topologies ($\Delta$ is a triangulation of the $3$-sphere
${S}^3$). Note that all edges are summed over, and therefore no reference is made to flat geometry.

Furthermore, we will show that this model is topological and that the weight associated to it can be interpreted algebraically as
 the $6j$-symbol of the Poincar\'e group.
The rest of the paper is devoted to the demonstration of these statements. The notations we use are summarized and explained in details in
appendix \ref{geo}.

\subsection{Emergence of the model}

\subsubsection{A key identity} \label{kid}

\begin{figure}[t]
\begin{center}
\includegraphics[width=3cm]{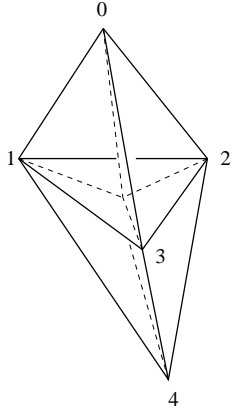}
\end{center}
\caption{Pentagonal representation of a complex of two tetrahedra $0,4$ sharing a face $\left[123\right]$, and a complex of
three tetrahedra $1,2,3$ sharing an edge $(04)$.} \label{pent}
\end{figure}

The keystone of the result announced above is a geometric identity associated with the possibility to embed a $4$-simplex
in $\mathbb{R}^3$.
A similar identity appears in \cite{Kor,Wood} as a basis to construct 3D manifold invariants.
Let $G\equiv\mathcal{V}^2$ be the square of the volume of a $4$-simplex $(0,\ldots,4)$ with edge lengths $l_{ij}$.
We denote by $\tau_j$ the tetrahedron obtained by dropping the point $j$ of the $4$-simplex, and by $\mathcal{V}_j$ its
volume. We will consider both the complex of the three tetrahedra $\tau_1, \tau_2$ and $\tau_3$ surrounding the edge $(04)$,
and the complex of the two tetrahedra $\tau_0, \tau_4$ sharing the face $\left[123\right]$ (see Fig \ref{pent}).
The orientation of $\tau_j$ is denoted by $\epsilon_j$.

We work with fixed lengths $l_{ij}, (ij) \not= (04)$, the length $l_{04}$ being free to fluctuate.
Then, the following identity of measures holds:
\beq \label{id3d}
4l_{04} \delta(G) = \sum_{\epsilon_0,\,\epsilon_4}\,\frac{\delta(l_{04} - l_{04}^{\epsilon_0\epsilon_4})}
{\mathcal{V}_0\,\mathcal{V}_4} = \sum_{\epsilon_1,\,\epsilon_2,\,\epsilon_3}
l^2_{04}\,\frac{\delta(\omega^{\epsilon}_{04}(l_{04}))}{\mathcal{V}_1\,\mathcal{V}_2\,\mathcal{V}_3}.
\eeq
The delta function in the third term is the $2\pi$-periodic delta function. The quantity $\omega^{\epsilon}_{04}$, considered as a function of $l_{04}$
and orientations, is the deficit angle carried by the edge $(04)$, defined to be:
\[
\omega^{\epsilon}_{04} = \sum_{j=1}^3\epsilon_j\,\theta_{04}^j
\]
where $\theta_{04}^j \in [0,\pi]$ is the interior dihedral angle of the edge $(04)$ in $\tau_j$. It represents the
curvature localized on the edge, in the sense of Regge calculus. It vanishes modulo\footnote{\label{explicit} If we don't work modulo
$2\pi$ the condition for the existence of an embedding of the  $4$-simplex $(0,\ldots,4)$ in  flat space is $\omega^{\epsilon}_{04} =  2\pi n(\epsilon)$
where $n(\epsilon)\in\{0,+1,-1\}$ depends on the orientations in this way:
\beq
n(\epsilon)= \begin{cases} \epsilon_1 \quad \text{if} \quad \epsilon_i=\epsilon_j \quad \forall \, (i,j) \\
\, 0 \qquad \text{if not}
\end{cases}
\eeq}
$2\pi$ if and only if the complex of three tetrahedra $(\tau_1, \tau_2, \tau_3)$ can be mapped in $\RR^3$ with the orientations $\epsilon_j$.
Thus, the delta function of the deficit angle acts as a projector on the space of flat geometries.

On the other hand, the complex of the two tetrahedra $\tau_0, \tau_4$ can always be mapped in $\RR^3$, in a unique
(modulo translations and rotations) way depending on the orientations.
The `missing' distance between the points $0,4$ is then a well defined function $l_{04}^{\epsilon_{0}\epsilon_4}$
of $l_{ij}$ and the product orientations.

\medskip

We now give a proof of (\ref{id3d}). We will need three intermediate results:
the first one states that the equation
\[
G(l_{04})=0,
\]
where $G \equiv\mathcal{V}^2$ is considered to be a function of $l_{04}$, all the other lengths being fixed, admits exactly two solutions $l_{04}^\pm$.
It was implicit in our discussion above, where we have mentioned, moreover, that these solutions
are labelled precisely by the product
$\epsilon_0\epsilon_4$. This result is geometrically clear since for each value of $l_{04}$ one can embed the simplex in $\RR^4$
in such a way to map, say, $\tau_0$ into a fixed tetrahedron. The volume of the simplex vanishes only
when the vertex $0$ belongs to the space $E$ spanned by the
tetrahedron $\tau_0$. As $l_{04}$ varies within its domain, the vertex $0$
moves along a circle around the face $\left[123\right]$, intersecting $E$ at
exactly two points.

We can also give an algebraic justification
for having only two solutions. Indeed, $G$ is the Cayley-Menger determinant of the 4-simplex.
The Cayley-Menger matrix is a $6\times 6$ symmetric matrix such that its diagonal elements are $0$ and the off diagonal
elements are  given by $L_{ab}\equiv -l^2_{ab}/2$, except for the first column and row where they are $1$.
Its determinant is therefore a quadratic function in $L_{04}$. Moreover, the coefficients of this polynomial can be
computed explicitly (see (\ref{derV}) and (\ref{derV2})):
\be
\frac{\partial G}{\partial L_{04}}= -2 \VV_{0}\VV_{4} \cos \theta_{04},
\quad
\frac{\partial^2 G}{\partial L_{04}^2}= -{2} \VV_{04}^2,
\ee
where $\theta_{04}$ is the interior dihedral angle between the two tetrahedra $0$ and $4$,
and $\VV_{04}$ is the area of the face $\left[123\right]$.

The second result states that the equation
\[
\omega_{04}^{\epsilon}(l_{04}) = 0,
\]
where $\omega$ is considered as a function of $l_{04}$ (the other lengths being fixed)
and orientations $\epsilon_{1}, \epsilon_{2}, \epsilon_{3}$, admits exactly four
solutions. Indeed this equation expresses the condition of existence of a map of the $4$-simplex in $\RR^3$ giving orientations
$\epsilon = \epsilon_1, \epsilon_2, \epsilon_3$
to the corresponding tetrahedra. Hence, first, it imposes $G=0$, and therefore, according to the previous result  $l_{04}=l_{04}^\pm$.
Now the data $\{l_{ij}, l_{04}^\pm\}$ determines the geometry of the complex in $\RR^3$ modulo orientation reversal in this space (in other words
simultaneous change of all orientations).
Therefore for each $l_{04}^\pm$, there exist exactly two sets of orientations $\eta \epsilon_i^\pm,\, \eta\in\{\pm1\}$ for which the equation is satisfied,
which proves the statement.

Let $(l_{04}^\pm,\epsilon_i)$ denote one of the four solutions of $G=\omega$=0. Then a variation $\delta l_{04}$ of the length
induces variations of $\delta G$ of the volume and
$\delta\omega$ of the deficit angle. Our third lemma, established in appendix \ref{geo}, relates these variations to each other:
\beqa \label{lemm1}
\left.\frac{\pa{G}}{\pa{\omega^{\epsilon}_{04}}}\right|_{l_{04}^\pm} &=&
-2\,\epsilon_1\epsilon_2\epsilon_3\frac{\mathcal{V}_1\mathcal{V}_2\mathcal{V}_3}{l_{04}}\\
\label{lemm2}
\left.\frac{\pa{G}}{\pa{l_{04}}}\right|_{l_{04}^\pm} &=&
\mp2\,\mathcal{V}_0\mathcal{V}_4l_{04}
\eeqa

The identity (\ref{id3d}) is obtained by evaluating in two different ways the
functional $\delta(G)$. We first consider $G$ to be a function of  $l_{04}$. We have:
\be\label{first}
\delta(G)=
\sum_{\epsilon\in\{\pm 1\}}\delta(l_{04}-l_{04}^{\epsilon})\left|\frac{\pa{G}}{\pa{l_{04}}}\right|^{-1}=
\sum_{\epsilon_0,\epsilon_4}\frac{1}{4l_{04}} \frac{1}{\VV_{0}\VV_{4}}\,        \delta(l_{04}-l_{04}^{\epsilon_0\epsilon_4}).
\ee
In order to get the second equality we have used (\ref{lemm2}), replaced the sum over the solutions by a sum over the
orientations of the tetrahedra $0,4$, taking into account the overcounting by a factor $2$.

Instead of considering $G$ as a function of $l_{04}$, one can consider it to be a function which depends on the
orientations $\epsilon_{1}, \epsilon_{2}, \epsilon_{3}$ and the corresponding deficit angle  $\omega^\epsilon_{04}$.
The following property
\[ G = 0 \, \Longleftrightarrow \, \exists \,\epsilon_{1}, \epsilon_{2}, \epsilon_{3} \quad {\rm such\, that}
\quad \omega^{\epsilon}_{04}=0 \quad  {\rm mod}\, 2\pi.
\]
yields:
\beq \label{last}
\delta(G) =\frac12\sum_{\epsilon_{1}, \epsilon_{2}, \epsilon_{3} }\delta(\omega^{\epsilon}_{04}) \left|\frac{\pa{G}}{\pa{\omega^{\epsilon}_{04}}}\right|^{-1}
          =\sum_{\epsilon_1,\,\epsilon_2,\,\epsilon_3}\frac{l_{04}}{4} \frac{\delta(\omega^{\epsilon}_{04})}{\mathcal{V}_1\,\mathcal{V}_2\,\mathcal{V}_3}
\eeq
The factor $1/2$ in the first equality comes from the fact that, if $\omega^{\epsilon}_{04}=0$, then $\omega^{-\epsilon}_{04}=0$,  so when we sum over all
$\epsilon_{i}$ we are overcounting with a factor of $2$. The equality between (\ref{first}) and (\ref{last}) leads to the identity (\ref{id3d}).

It is worth noting that one can actually derive an identity slightly stronger that (\ref{id3d}). In fact,
(\ref{lemm2}) divided by (\ref{lemm1}) gives the derivative of the deficit angle with respect to the length:
\beq \label{der}
\left.\frac{\pa \omega^{\epsilon}_{04}}{\pa l_{04}}\right|_{l_{04}^\pm} = \pm \epsilon_1\epsilon_2\epsilon_3 l_{04}^2 \frac{\VV_0\VV_4}{\VV_1\VV_2\VV_3}
\eeq
It is then straightforward to check that
\beqa
\forall \, \eta, \quad
\frac{f(l_{04}^\pm)}{\VV_0\VV_4} &=& \int_{l_{04}^{\pm}} \extd l_{04}l^2_{04}\,\frac{\delta(l_{04} - l_{04}^\pm)}{\mathcal{V}_1\,\mathcal{V}_2\,\mathcal{V}_3}
\left|\frac{\pa \omega^{\eta \epsilon^{\pm}}_{04}}{\pa l_{04}}\right|^{-1} f(l_{04})\\
\label{idbis}
&=& \int_{l_{04}^\pm} \extd l_{04}l^2_{04}\ \frac{\delta(\omega^{\eta \epsilon^\pm}_{04})}{\VV_1\VV_2\VV_3}f(l_{04}).
\eeqa
where the label appearing as an index of integration means that on integrates in the neighborhood of the value $l_{04}^\pm$.



\medskip

In the next two parts of this section, we will take advantage of
this identity to express the Feynman graph amplitude as a spin
foam amplitude.

\subsubsection{Spin foam in Feynman graph: inducing the geometry dynamically}

In order to understand the mechanism, let us begin by studying a
simple example. Suppose that our Feynman graph $\Gamma$ has the
structure depicted in Fig \ref{pent}: it admits five vertices and
ten links. The two tetrahedra $\tau_0, \tau_4$ form a triangulation
of a $3$-ball. According to the construction described in section
\ref{inv}, the amplitude of this graph is: \beq \label{ex1}
I_{\Gamma} = \int \prod_{(ij)\not=(04)}l_{ij}\mathrm{d}l_{ij}
\sum_{\epsilon_0,\, \epsilon_4} \frac{1}{\mathcal{V}_0\mathcal{V}_4}
\mathcal{O}_{\Gamma}(l_{ij}, l^{\epsilon_0\epsilon_4}_{04}(l_{ij}))
\eeq where $\mathcal{O}_{\Gamma}$ is the product of propagators
\[
G^F(l^{\epsilon_0\epsilon_4}_{04}(l_{ij}))\prod_{(ij)\not=(04)}G^F(l_{ij})
\]
 The length of $(04)$ depends explicitly on the nine other lengths.
The previous identity (\ref{id3d}) allows us to remove this dependence by introducing a new label $l_{04}$ and summing over it.
Plugging it into (\ref{ex1}), and expanding the periodic delta function, as a sum $2 \pi\delta(\omega)=\sum_{s \in \ZZ} e^{\imath s \omega}$, we get:
\beq \label{ex2}
I_{\Gamma}=\frac{1}{2\pi}\int\prod_{(ij)\neq (04)}l_{ij}\mathrm{d}l_{ij}\int \mathrm{d}l_{04}l^2_{04}\sum_{s_{04}\in \ZZ}
\sum_{\epsilon_1,\,\epsilon_2,\,\epsilon_3}
\frac{e^{\imath s_{04}\omega^{\epsilon}_{04}} }{\mathcal{V}_1\mathcal{V}_2\mathcal{V}_3} \mathcal{O}_{\Gamma}(l_{ij}, l_{04}) \,
\eeq
Notice that the tetrahedra $\tau_1,\tau_2,\tau_3$ form a new triangulation of the ball. In this expression, one begins to recognize the structure of the
model (\ref{id3d}), since the total action can be expressed in terms of the deficit angles associated with the edges of $\Delta$
\be
S_{\Delta}\equiv \sum_{\tau} \epsilon_{\tau} S_{\tau} = \sum_{e} s_{e} \omega^{\epsilon}_{e}.
\ee

Let us summarize where we stand so far in addressing the first step, which was
to transfer the information about flat geometry from the measure into relations
$l^{\epsilon}_{e'}(l_e)$; it has led us from (\ref{intx}) to (\ref{Feyamp}). Now considering two tetrahedra sharing a
face in the triangulation $\Delta_k$, we have learned from the example above that the `missing' distance variable can be replaced by a free label
that we integrate over. The price to pay
is the apparition of a constraint term - the delta function of the deficit angle - which, in the language of Regge calculus, imposes locally,
the geometry to be flat. Ultimately, by expanding the delta function, we begin to see a model emerging from the Feynman graph,
where the flat geometry is induced dynamically.

The result (\ref{exp}) will be proved in two steps. First, we will to interpret our key identity as expressing the invariance under Pachner moves,
and thus establish the topological invariance of the model (\ref{3dmod}).
Secondly we will  write the measure of the Feynman graph as the partition function of this model, computed on a specific triangulation.

\subsection{Topological invariance}

\label{topo}

In this section we  show that the spin foam model (\ref{3dmod}),
\beq \label{3dmod'}
Z_{\Delta}=\frac{1}{(2\pi)^{|e|}}\int_{\mathrm{\rm{GF}}}\prod_{e\in\Delta}\mathrm{d}l_el^2_e\sum_{\{s_e,\epsilon_{\tau}\}}\left(\prod_{\tau}\,
\frac{e^{\imath \epsilon_{\tau} S_{\tau}(s_{e},l_{e})}}{\mathcal{V}_{\tau}}\right),
\eeq
is independent of the choice of the triangulation and depends only on the topology of the manifold that $\Delta$ triangulates.
The label (GF) appearing as an
index of integration means that the integrals have to be \textit{gauge fixed}, in order to properly define the partition function.

The necessity of gauge fixing the model is manifest if one carefully looks at the integrand of (\ref{3dmod}). In fact,
the sum over integers $s$ of the exponentials leads to a product over edges of delta functions $\delta(\omega^{\epsilon}_e)$.
The constraints $\omega^{\epsilon}_e =0 \, \mod 2\pi$ are not all independent however, and the redundancy of delta functions induces divergences.
These naive divergences are related to the existence of gauge symmetries in the model, which have to be gauge fixed.
This is analogous to what happens in the study of the 3d quantum gravity amplitudes, as explained in \cite{LDA2}.
The symmetries and the corresponding gauge fixing of the model will be studied in great detail in the next section .

In order to specify this gauge fixing, we first choose $4$ vertices  which belong to a tetrahedron $\tau_{0}$ of $\Delta$, and then assign to any other
internal vertex $v$ of $\Delta$ a tetrahedron $\tau_{v}$ to which this vertex belongs. This assignment is said to be admissible if
the following condition is met: given two vertices $v_{1}$ and $v_{2}$ of $\Delta$ that are not in $\tau_{0}$, then
$(v_{1}v_{2}) \not\subset \tau_{v_{1}}\cap \tau_{v_{2}}$, that is, the edge $(v_{1}v_{2})$ (when it exists) does not belong to the
intersection of the corresponding
tetrahedra. We denote such an admissible gauge fixing assignment $\left(\tau_{0}, \{\tau_{v}\}_{v\not\in \tau_{0}}\right)$ by $T$.
We say that a tetrahedron $\tau$ belongs to $T$ if either $\tau =\tau_{0}$ or $\tau=\tau_{v}$ for some vertex not in $\tau_{0}$;
we say that an edge $e$ belongs to $T$ if either $e\in \tau_{0}$ or $e \in \tau_{v}$, and if $e$ admits $v$ as one of its vertex -
for each $\tau_{v}$ there are three such edges, carrying the labels $l_{v1}, l_{v2}, l_{v3}$.
The admissibility condition insures that these triplets of edges are independent, that is, we cannot have $l_{vi}=l_{v'j}$.
We will see later that the admissibility requirement can always be met.

We now define a gauge fixing term $\delta_{GF}^T$ and a Faddeev-Popov determinant $D_{FP}^T$ as
\beqa \label{gfterm}
\delta_{\mathrm{GF}}^T(l_{e},s_{e})&=& \prod_{e\in T}\delta(l_{e} - l_{e}^{o}) \, (2\pi) \delta_{s_{e},s_{e}^{o}} \\
D_{\mathrm{FP}}^{T}(l_{e},s_{e}) &=& \prod_{\tau \in T} \frac12 \left(\frac{ \VV_{\tau}}{l_{v1}l_{v2}l_{v3}}\right)^{2}.
\eeqa
The gauge fixing term imposes the values of $l_{e}$ and $s_{e}$ to be
$l_{e}^{o}, s_{e}^{o}$, which are arbitrary fixed. Fixing the labels $s$'s living on the edges of the assignment removes the redundant delta functions
of the deficit angle at these edges. Now note that the constraints $\omega_e^\epsilon(l) = 0 \left[2\pi\right]$
act both on the labels $l$'s and the orientations. Therefore by removing them, we not only remove redundant constraints on the $l$'s but also
necessary restrictions acting on orientations. As shown in appendix, one can reintroduce these restrictions by inserting an additional gauge fixing
factor\footnote{This factor acts `on-shell' as Kronecker symbols for the orientations and is understood here as a gauge fixing. An interesting open question is
whether this constraint can be expressed off-shell and implemented dynamically.} in $\delta_{GF}^T$:
$$ \prod_{v\in T} \chi(\Omega_v^{\epsilon}(l_e)).$$
The function $\chi(x)$ is a characteristic function defined to be constant, with value $1$, on
$4\pi\mathbb{Z}$, and $0$ elsewhere; and $\Omega_v^{\epsilon} = \sum_{\tau \supset v} \epsilon_{\tau} A_{\tau}$ is
the algebraic sum, over all tetrahedra sharing the vertex $v$, of the solid angle $A_{\tau}$ at $v$ within $\tau$.

The gauge fixed partition function is obtained by inserting in the integral (\ref{3dmod'}) the product $\delta^T_{\mathrm{GF}}D^T_{\mathrm{FP}}$.

\medskip

Showing the topological invariance of the model amounts to showing the invariance of $Z_{\Delta}$ under the Pachner moves $(2,3)$ and $(1,4)$,
and its independence of the choice of $T$. The invariance under the move $(2,3)$ follows from the pentagonal identity,
\be \label{2,3}
\sum_{\epsilon_0, \epsilon_4}
\frac{e^{\imath \epsilon_{0} S_{0}}}{\mathcal{V}_0} \frac{e^{\imath \epsilon_{4} S_{4}}}{\mathcal{V}_4} =
\frac{1}{2\pi}\,\sum_{\epsilon_{1},\epsilon_{2},\epsilon_{3}}
\int \mathrm{d}l_{04}\,l^2_{04} \sum_{s_{04}}
\frac{e^{\imath \epsilon_{1} S_{1}}}{\mathcal{V}_1}
\frac{e^{\imath \epsilon_{2} S_{2}}}{\mathcal{V}_2}
\frac{e^{\imath \epsilon_{3} S_{3}}}{\mathcal{V}_3},
\ee
where $\tau_{0}, \cdots \tau_{4}$ labels the five tetrahedra which triangulate the boundary of a $4$-simplex,
$\tau_{i}$ being the tetrahedron where the vertex $i$ is omitted; the invariance under the move $(1,4)$ is due to the $(1,4)$ identity
\beq \label{1,4}
\sum_{\epsilon_0}\frac{e^{\imath \epsilon_{0} S_{0}}}{\mathcal{V}_0} =
\frac{1}{(2\pi)^4}\sum_{\epsilon_{j}}
\int\prod_{j=1}^{4}\mathrm{d}l_{0j}l_{0j}^2 \sum_{s_{0j}}\, \prod_{j=1}^{4}\frac{e^{\imath \epsilon_{j} S_{j}}}{\mathcal{V}_j}\delta_{\mathrm{GF}}D_{\mathrm{FP}}
\eeq
$\delta_{\mathrm{GF}}$ denotes the gauge fixing term and  $D_{\mathrm{FP}}$ the Faddeev-Popov determinant:
\be \label{gfterms}
\delta_{\mathrm{GF}}= (2\pi)^3\prod_{i=1}^3\delta(l_{0i}-l_{0i}^o)\delta_{s_{0i}, s_{0i}^o} \chi(\Omega_0^{\epsilon}),
\qquad
D_{\mathrm{FP}}=  \frac{\mathcal{V}_4^2}{2\prod_{i=1}^3l_{0i}^2}
\ee
where $l_{0i}^{o}, s_{0i}^o$ are any fixed values.
At this stage the insertion of these terms are needed for topological invariance;
they will be fully and independently justified in the next section (together with the gauge fixing denomination).

\medskip
The proof of (\ref{2,3}) is a simple adaptation of the argument leading to the key identity (\ref{id3d}).
As before, we denote by $\eta \epsilon_i^{\pm}, \eta = \pm1$, the two sets of orientations for which the deficit angle
$\omega_{04}^{\epsilon}(l_{04}^{\pm})$ vanishes. If this dihedral angle is zero modulo $2\pi$, the $4$-simplex $[01234]$ can be mapped in $\RR^3$.
Each edge $(ab)$ belongs then to a complex of three tetrahedra
embedded in $\RR^3$. One can therefore find orientations $\epsilon_0^\pm, \epsilon_4^\pm$ of  $\tau_0,\tau_4$, for which all deficit angles
$\omega^{\epsilon_{\pm}}_{ab}$ are also zero modulo $2\pi$.
The actions of the $(2,3)$ move evaluated for $l_{04}=l_{04}^\pm$ are then related by
\be \label{rel}
\epsilon_{0}^{\pm} S_{0} +\epsilon_{4}^{\pm}S_{4} + \epsilon_1^{\pm}S_{1}^{o}+\epsilon_2^{\pm}S_{2}^{o}+\epsilon_3^{\pm} S_{3}^{o}
\equiv \sum_{(ab)\neq (04)} s_{ab} \omega_{ab}^{\epsilon^\pm} = 0 \,\left[2\pi\right]
\ee
The superscript $o$ means that $s_{04}=0$ in the actions. This equality, together with (\ref{idbis}), show that
\be \label{idbase}
\forall \, \eta, \quad
\frac{e^{\imath \eta \epsilon^{\pm}_{0} S_{0}}}{\mathcal{V}_0}
\frac{e^{\imath \eta \epsilon^{\pm}_{4} S_{4}}}{\mathcal{V}_4} =
\int_{l_{04}^{\pm}} \extd l_{04} l^2_{04} \delta(\omega_{04}^{\eta \epsilon^\pm})
\prod_{i=1}^3\frac{e^{-\imath \eta \epsilon_{i}^\pm S_{i}^{o}}}{\mathcal{V}_i}
\ee
Now we want to sum over $\eta = \pm1$. Since $\eta \epsilon_i^\pm$ are the only possible choice
of orientations for  which
$\omega^{\epsilon}_{04}= 0$ admits a solution around $l_{04}^\pm$, one can replace, in the RHS,  the summation over $\eta$ by a sum over all
orientations. Then summing the contributions of the two values $l_{04}^\pm$ leads to
\be
\sum_{\pm}
\sum_{\eta}
\frac{e^{\imath \eta \epsilon^{\pm}_{0} S_{0}}}{\mathcal{V}_0}
\frac{e^{\imath \eta \epsilon^{\pm}_{4} S_{4}}}{\mathcal{V}_4} =
\sum_{\epsilon_{1},\epsilon_{2},\epsilon_{3}} \int \mathrm{d}l_{04}\,l^2_{04} \delta(\omega_{04}^{\epsilon})
\prod_{i=1}^3\frac{e^{\imath \epsilon_{i} S_{i}^{o}}}{\mathcal{V}_i} .
\ee
Now it is easy to check\footnote{In fact, our definitions of orientations (see appendix) imply that $\epsilon_0^\pm\epsilon_4^\pm = \pm1$.
It shows that
the application $(\pm,\eta) \mapsto (\eta\epsilon_0^\pm, \eta\epsilon_4^\pm)$ is surjective onto $\{-1, +1\}$.}
that the summation in the LHS can be replaced by a sum over the values of orientations $\epsilon_0, \epsilon_4$.
If one now  expands the delta function in the RHS as a
sum over $s_{04}$ of $1/ 2\pi  \exp{(\imath s_{04}\omega_{04})}$, one obtains  (\ref{2,3}). It's worth noting that
we could have  inserted any function $f(l_{04})$
of the length $l_{04}$ in both sides of all our equalities, without modification, and therefore $(\ref{2,3})$ should be understood as an equality of
measures on the space of such functions.
Since $S_{4}$ and $\VV_{4}$ do not depend on $l_{04}$, we multiply both sides of (\ref{idbase}) by
$\VV_{4}e^{-\imath \eta\epsilon_{4}^\pm S_{4}}$, to get
\be \forall \, \eta, \quad
\frac{e^{\imath \eta \epsilon^{\pm}_{0} S_{0}}}{\mathcal{V}_0} =
\int_{l_{04}^{\pm}} \mathrm{d}l_{04}\,l^2_{04}
\delta(\omega_{04}^{\eta \epsilon^\pm})
\prod_{i=1}^3\frac{e^{- \imath \eta \epsilon_{i}^\pm S_{i}^{o}}}{\mathcal{V}_i}
 \frac{e^{-\imath \eta\epsilon_{4}^\pm S_{4}}}{\mathcal{V}_4} \VV_{4}^2,
\ee
where in the RHS the values of $l_{4i}\equiv l_{4i}^{o}$ are fixed
to be any value - since the reasoning leading to (\ref{idbase}) does not depend on it.
We now insert the trivial `gauge fixing' identities
\beqa
1 &=& \int \prod_{i=1}^3 dl_{0i} l_{0i}^2  \prod_{i=1}^3 \frac{\delta(l_{0i}-l_{0i}^o)}{l_{0i}^2}\\ \label{gfix2}
1 &=& \sum_{\epsilon_4} \frac{1}{(2\pi)^3}\sum_{s_{0i}} \prod_{i=1}^3 \delta_{s_{0i}, s_{0i}^o}(2\pi)^3 \chi(\Omega_v^{\epsilon}),
\eeqa
where $\epsilon = (\eta \epsilon_i^\pm, \epsilon_4)$.  As can be easily seen geometrically, and as more precisely shown in appendix,
the algebraic solid angle $\Omega_0^{\epsilon}$, evaluated at $l_{04}^\pm$,
vanishes (modulo $4\pi$) only if $\epsilon_4 = \eta\epsilon_4^\pm$, and therefore its image by the function $\chi$ acts as a Kronecker
symbol for $\epsilon_4$.
Hence, we obtain
\be
\frac12 \frac{e^{\imath \eta \epsilon^{\pm}_{0} S_{0}}}{\mathcal{V}_0} =
\frac{1}{(2\pi)^4} \sum_{\epsilon_4}
\int \prod_{i=1}^3 \mathrm{d}l_{0i} \,l^2_{0i} \int_{l_{04}^{\pm}} \mathrm{d}l_{04} \,l^2_{04}
\sum_{s_{0i}} \delta(\omega_{04}^{\eta \epsilon^\pm})
\prod_{i=1}^3\frac{e^{- \imath \eta \epsilon_{i}^\pm S_{i}^{o}}}{\mathcal{V}_i}
 \frac{e^{-\imath \epsilon_{4} S_{4}}}{\VV_4} \delta_{GF}D_{FP}.
\ee
for all $\eta=\pm1$. Eventually,
summing as before both sides of this equality over $\eta$, adding the contributions of $l_{04}^+$ and $l_{04}^-$,
and expanding  the delta function in the RHS, show (\ref{1,4}).

\subsection{Symmetries and gauge fixing} \label{sym3d}

We have seen in the previous part that, in order to have a model which is topologically invariant, we need to
insert gauge fixing and Faddeev-Popov terms in the definition of the spin foam model (\ref{3dmod}).
We now want to show that these factors are not only needed for topological invariance, but arise naturally from the gauge fixing of the
symmetries of the action.

\medskip

The action that arises from our construction reads:
\beq  \label{ac3d}
{S}_{\Delta}[s_e,l_e]\,=\sum_{\tau}\epsilon_{\tau} S_{\tau}(s_e,l_e)\,=\sum_e\,s_e\,\omega_e^\epsilon(l_e)
\eeq
The naive divergences
of the model (\ref{3dmod}) can be seen as arising from symmetries of the model mapping
solutions to solutions. Fixing these symmetries, treated as gauge
 symmetries, will lead to the Faddeev-Popov determinants
described in the previous section.
In order to characterize these symmetries infinitesimally, we will study the zero modes of the Hessian of
${S}_{\Delta}$. Here $s_{e}$ will be treated as a continuous variable, in order to simplify the analysis\footnote{We can treat $s_{e}$
as a continuous variable if one authorizes $\omega_{e}$ to be shifted by $2n\pi$ for $n$ an arbitrary integer.
This is clear from the expression of the periodic delta function $\delta(\omega) =\sum_{n} \int ds e^{\imath s (\omega-2n\pi)}$.
This affects the classical solutions but not the equations that involve derivatives of $\omega$, which are our main interest here.}.

Let us first describe the classical solutions. The equations of motion (eom) are
\beqa \label{eom}
\label{cl1} 0 &=& \frac{\delta S}{\delta s_e}=\omega^{\epsilon}_e(l) \quad \forall e \\
\label{cl2} 0 &=& \frac{\delta S}{\delta l_e}= \sum_{\tau}\epsilon_{\tau}\left(\sum_{e'\in\tau}
s_{e'}\frac{\delta \theta^{\tau}_{e'}}{\delta l_e}\right)= \sum_{e'}s_{e'}\frac{\delta \omega^{\epsilon}_{e'}}{\delta l_{e}}
\quad \forall e
\eeqa
The first equation expresses the flatness
condition. A set of lengths $\{l_e^0\}$ is solution if
the complex can be locally\footnote{A global map exists, and is unique modulo translations and rotations, if the manifold that $\Delta$ triangulates
is simply connected \cite{Wood}.} mapped in $\RR^3$ in a way that respects orientations $\epsilon$.
The second equation is reminiscent of the \textit{Schl\"{a}fli identity} for flat tetrahedra
\beq
\sum_{e'\in\tau}
l_{e'}\frac{\delta \theta^{\tau}_{e'}}{\delta l_e}=0 \,\, \to \,\, \sum_{e'}l_{e'}\frac{\delta \omega^{\epsilon}_{e'}}{\delta l_{e}}=0.
 \eeq
This shows that a solution of (\ref{cl2}) is given by the values $s^0_e= \alpha l_e$, with $\alpha$ an arbitrary constant.
Notice that if one inserts this solution into the action (\ref{ac3d}), one recovers the
Regge action of discrete $3D$ gravity
\beq \label{re}
\mathcal{S}_R=\alpha \sum_el_e\omega^{\epsilon}_e(l_e)
\eeq
In the following a solution $\{l^0_e,s^0_e=l^0_e\}$ of the eom will be called a Regge solution (we put the overall scale factor $\alpha=1$ for simplicity).

Zeros modes are infinitesimal transformations of the fields (here
the labels) that belong to the kernel of the Hessian
$\delta^2{S}$, computed on shell.  This kernel is characterized by
deformations $\delta s_e, \delta l_e$ that satisfy the equations
\beqa
\label{h1}\sum_{e'}\,\frac{\delta\omega_e}{\delta
l_{e'}}\,\delta l_{e'}\,&=&\,0 \quad \forall e\\
\label{h2}
\sum_{e',e''}\,s_{e''}\,\frac{\delta^2\omega_{e''}}{\delta
l_e\delta l_{e'}}\,\delta
l_{e'}\,+\,\sum_{e'}\,\frac{\delta\omega_{e'}}{\delta l_e}\,\delta
s_{e'}\,&=&\,0 \quad \forall e
\eeqa
where the label $\epsilon$ has been dropped for clarity.
This system of equations can be rearranged if one uses the derivative of the Schl\"{a}fli identity
\beq
 \frac{\delta}{\delta l_e}\,\left[\sum_{e''}l_{e''}\,\frac{\delta\omega_{e''}}{\delta
l_{e'}}\right]
= \frac{\delta\omega_e}{\delta
l_{e'}}\,+\,\sum_{e''}l_{e''}\frac{\delta^2\omega_{e''}}{\delta
l_e\delta l_{e'}} =0,
\eeq
and the fact that the matrix
$\left(\frac{\delta \omega_e}{\delta l_{e'}} \right)$ is symmetric,
as the Hessian of the Regge fonction (\ref{re}).
We can write (\ref{h2}) as
\be
\sum_{e',e''}\,(s_{e''}-l_{e''})\,\frac{\delta^2\omega_{e''}}{\delta l_e\delta l_{e'}}\,\delta
l_{e'}\,+\,\sum_{e'}\,\frac{\delta\omega_{e}}{\delta l_{e'}}\,
\delta(s_{e'}-l_{e'})\,=\,0.
\ee
If one focuses on fluctuations around Regge solutions, the system becomes equivalent to
\beq \label{sys}
\sum_{e'}\,\frac{\delta\omega_e}{\delta l_{e'}}\,\delta l_{e'}\,
=\,\sum_{e'}\,\frac{\delta\omega_e}{\delta l_{e'}}\,\delta s_{e'}\,=\,0 \qquad \forall e
\eeq

One can easily find infinitesimal variations  $\delta l_{e}, \delta s_{e}$
solutions of (\ref{sys}). Indeed, an embedding in $\RR^3$ of the complex $\Delta$,
with the lengths $\{l^0_e\}$, gives positions to the vertices.
Infinitesimal moves of the vertices in $\RR^3$ around these positions induces
deformations of the edge lengths. By construction, the new deficit
angles still vanish. Thus, these deformations correspond to one of the
symmetries we are considering, acting on vertices of the
triangulation. The positions of the vertices in an embedding are
therefore pure gauge in our model. The generators are 3-vectors $\vec{\alpha}_v$ attached
to each vertex $v$ which induce displacements of $v$ in
$\RR^3$. They determine variations
$\delta s_{e} =\delta l_{e}(\vec{\alpha}_{v},l_{e}^0) $,  $\delta l_{e} =\delta l_{e}(\vec{\beta}_{v},l_{e}^0) $,
which are solutions to the system of equations (\ref{sys}).

These transformations map solutions to solutions\footnote{\label{res} These symmetries, acting on the space of solutions, do not however
leave the full action invariant away from classical configurations.
Whether or not these symmetries admit an off-shell extension is a interesting open question not addressed here.};
the corresponding symmetry group
is $(\RR^3 \times \RR^3)^{|v|}$, where $|v|$ is the number of vertices of the triangulation.
The gauge fixing is performed by fixing a subset of edge lengths while taking into account the Jacobian or Faddeev-Popov determinant.
These Jacobians can easily be read out from the results of  section \ref{inv}.
We start by choosing four vertices that form a tetrahedron $\tau_{0}$ in the triangulation.
For these vertices we have
\be
\extd^3 \alpha_{1}\extd^3 \alpha_{2}\extd^3 \alpha_{3}\extd^3\alpha_{4} = 2\extd^3 \alpha \extd \Lambda
\frac{\prod_{i<j} l_{ij} \extd l_{ij}}{\VV_{\tau_{0}}}.
\ee
The factor $2$ is due to the sum over values of the orientation for $\tau_0$.
For each additional vertex $v$, we choose a tetrahedron $\tau_{v}$, and denote $l_{v_{1}},l_{v_{2}},l_{v_{3}}$ the three edges of $\tau_{v}$ touching $v$.
In this case we have
\beq \label{fac}
\extd^3 \alpha_{v}=  2\frac{\prod_{i} l_{v_{i}} \extd l_{v_{i}}}{\VV_{\tau_{v}}}.
\eeq
Each time we add a new vertex $v_{n}$, one should insure that $\tau_{v_{n}}$ is picked in such a way that the three corresponding edges
do not overlap with
any of the previously chosen edges  $l_{v_{1i}},\cdots l_{v_{n-1i}}$,
in order for an edge to be selected at most once in this process.
We are going to see in the following how this can always be concretely implemented for any closed triangulation.

This  procedure specifies an admissible gauge fixing  assignment $T_{l}$, as defined in the previous section, which allows us to gauge fix the symmetry
acting on $l$'s.
The gauge fixing term is given by
\be
\delta^{T_{l}}_{GF}= \prod_{e\in T_{l}} \delta(l_{e}-l_{e}^{o})
\ee
and the determinant by
\be \label{fpl}
(D^{T_{l}}_{FP})^{-1}\equiv \frac{\bigwedge_{v} \extd^3 \alpha_{v}}{\extd^3 \alpha \extd \Lambda \bigwedge_{e\in T_l} \extd l_{e}}
= 2^{|v|-3} \frac{\prod_{e\in T_l} l_{e}}{\prod_{\tau \in T_l} \VV_{\tau}},
\ee
where $|v|$ is the number of vertices of $\Delta$.
The same procedure is applied to the gauge fixing of the symmetry acting on $s$'s,
with a gauge fixing assignment $T_s$, and the corresponding gauge fixing term and determinant
\be
\delta^{T_{s}}_{GF} = \prod_{e\in T_{s}} (2\pi) \delta_{s_{e},s_{e}^{o}}\prod_{v \in T_s} \chi(\Omega_v^{\epsilon}(l_e))
\ee
\be
\label{fps}
(D^{T_{s}}_{FP})^{-1}\equiv \frac{\bigwedge_{v} \extd^3 \alpha_{v}}{\extd^3 \alpha \extd \Lambda \bigwedge_{e\in T^s} {\extd s_{e} } }
=  \frac{\prod_{e\in T^s}  l_{e}}{\prod_{\tau \in T^s} \VV_{\tau}}.
\ee
There is no factor $2$ in this Faddeev-Popov determinant, since here orientations of the $s$'s tetrahedra are fixed
thanks to the factors $\chi(\Omega)$ in the gauge fixing term.
The insertion of the functions $\chi$ restricts the values of orientations to those for which solutions of the eom (\ref{eom}),
around which the zeros modes which  have to be fixed, exist.
One thus recovers the gauge fixing described in the previous section when one chooses the same gauge fixing assignment $T_l=T_s$ for $l$ and $s$.

\medskip

Finally, we need to show that an admissible  gauge fixing assignment can always be reached.
One way to construct such an assignment is to choose a rooted tree (also denoted $T$ for simplicity) in the one skeleton of the dual triangulation.
The one skeleton of the dual triangulation is a $4$-valent graph;  each
vertex of this graph corresponds to a tetrahedron of $\Delta$. A tree is a connected subgraph of this skeleton which contains no loop.
A root is a choice of one particular end point of this tree diagram. We use the notation $v^{\star}$ for the vertices of the dual triangulation,
and $\tau_{v^\star}$ for the tetrahedron of $\Delta$ dual to $v^\star$. A tree $T$ is said \textit{maximal}
if
each vertex of $\Delta$ belongs to a tetrahedron dual to a vertex of $T$:
$\forall v \in \Delta, \exists v^\star \in T, v \subset \tau_{v^\star}$

Given a maximal rooted tree $T$ one can construct an admissible assignment as follows.
The root of $T$ corresponds to a choice of an initial tetrahedron $\tau_{0}$. Then, lets pick an edge $e^\star$
that goes from the root to a nearby tetrahedron $\tau_{e^\star}$: this edge is dual to a face $f_{e^\star}$ of the
triangulation. If we consider $\tau_{e^\star}-f_{e^\star}$, this consists of three edges meeting at a vertex $v_{e^\star}\in \tau_{e^\star}$; in this way
we define a correspondence between $e^\star$ and a pair $(\tau_{e^\star},v_{e^\star})$.
This correspondence can be extended to the all tree if one uses the natural orientation of the rooted tree.
Indeed, since the tree is rooted, every edge of the tree can be oriented with the orientation
always going farther away from the root. And to any edge $e^\star$ of $T$ we assign a tetrahedron $\tau_{e^\star}$ corresponding to the end
point of this edge and a vertex $v_{e^\star}$ being the unique vertex in $\tau_{e^\star}-f_{e^\star}$.
This assignment is clearly admissible. In order to see why, lets suppose that there exists
an edge of $\Delta$ that belongs to both $\tau_{e^\star}-f_{e^\star}$ and $\tau_{e'^\star}-f_{e'^\star}$.
This edge would belong to the intersection; but this intersection
is empty since tetrahedra always intersect on faces dual to the rooted tree.
Eventually, the maximality of the tree implies the maximality of the gauge fixing assignment.

\subsection{Observables and partial gauge fixing}

In this section we will show our main result, namely that the original Feynman integral (\ref{Feyamp})
can be expressed as an expectation value of an observable in our topological spin foam model.

\medskip

We have seen that the pure model (\ref{3dmod}) has to be gauge fixed; the complete gauge fixing is performed by choosing a admissible
assignment $T$,  and by inserting gauge fixing and Faddeev-Popov terms
$K^{T_s,T_l}(l_e, s_e) = \delta^{T_{s}}_{GF}\delta^{T_{l}}_{GF}D^{T_{s}}_{FP}D^{T_{l}}_{FP}$, with $T_s = T_l = T$.
We will call \textit{observable} of the model a function $f(l_e,s_e)$ of the labels, and define its evaluation as
\beq
\langle f \rangle_{\Delta} = \frac{1}{(2\pi)^{|e|}} \int_{GF} \prod_{e\in\Delta}\mathrm{d}l_el^2_e
\sum_{\{s_e,\epsilon_{\tau}\}} f(l_e,s_e)
\left(\prod_{\tau}\, \frac{e^{\imath \epsilon_{\tau} S_{\tau}(s_{e},l_{e})}}{\mathcal{V}_{\tau}}\right)
\eeq
$f$ is not in general gauge invariant and therefore its insertion modifies the discussion of the gauge fixing. We will be interested in particular in the
class of observables that are functions of the lengths only, of a subgraph $\Gamma$ drawn in $\Delta$. The typical example is the so-called
matter observable
$\OO(l_e) = \prod_{e\in{\Gamma}} G^F(l_e)$, where $\Gamma$ is a subgraph of the $1$-skeleton of $\Delta$.

The introduction of such observables breaks a part of the gauge
symmetries - namely the symmetry of the field $l_e$ which acts on the vertices of $\Gamma$, turning the former gauge degrees of freedom
into dynamical degrees of freedom.
In the presence of matter observables the gauge fixing can no longer be symmetrical, and one needs to choose different assignments
for the $l$'s and $s$'s gauge fixing. Let $T_s$ be a maximal admissible assignment $\{\tau_v\}$, and
$T_l \subset T_s$ a subset that is a maximal assignment in $\Delta\backslash \Gamma$. In other words, $T_l = \{\tau_v\}_{v \notin \Gamma}$.
We then take care of the remaining gauge symmetries  by inserting the
gauge fixing and Faddeev-Popov terms corresponding to these assignments. The evaluation of the observable $\OO$ in the (partially)
gauge fixed model reads then:
\beq \label{eval}
\langle \OO \rangle_{\Delta} = \frac{1}{(2\pi)^{|e|}} \int \prod_{e\in\Delta}\mathrm{d}l_el^2_e\sum_{\{s_e,\epsilon_{\tau}\}} \OO(l_e)
\left(\prod_{\tau}\, \frac{e^{\imath \epsilon_{\tau} S_{\tau}(s_{e},l_{e})}}{\mathcal{V}_{\tau}}\right) K^{T_s,T_l}(l_e,s_e)
\eeq

We now suppose that $\Gamma$ is a Feynman graph and $\Delta$ a triangulation of the $3$-sphere $\mathcal{S}^3$ such that
$\Gamma \subset \Delta$.
We want to show that the expectation value $(\ref{eval})$ is equal to the Feynman amplitude (\ref{Feyamp}) of the graph $\Gamma$.

Recall that the triangulation $\Delta_k$ involved in (\ref{Feyamp}) is a triangulation of a $3$-ball, without any internal vertex nor
internal edge; it is not a triangulation of $S^3$. It is easy however to construct a triangulation of
$S^3$ from  $\Delta_k$, simply by taking its \textit{double} denoted $D\Delta_k\equiv \Delta_k  \sharp_{S^2} \bar{\Delta}_k $,
built with a copy of $\Delta_k$ in the interior of a 3-ball $B_{3}\subset\mathcal{S}^3$ glued to
a copy of $\Delta_k$ in the exterior $\mathcal{S}^3 \backslash B_{3}$.

The proof proceeds in three steps. Primarily, since both $D\Delta_k$ and $\Delta$ are triangulations of $S^{3}$ which
contain vertices of the graph $\Gamma$, the two triangulations can be constructed out of each other by a sequence of
Pachner moves that do not remove the vertices of $\Gamma$.
Secondly, $\langle \OO \rangle_{\Delta}$ is invariant under these moves, provided that the variable $l_e$ of each edge $e \in \Gamma$
that is removed by a $(2,3)$ move is replaced by its value `on shell', namely the one fixed by the constraint $\omega_e^{\epsilon}=0$.
The justification comes directly from the analysis of section \ref{topo}.
One can therefore write
\beq
\langle \OO \rangle_{\Delta} = \langle \tilde{O}_{\Gamma} \rangle_{D\Delta_k}, \quad \text{with} \quad
\tilde{O}_{\Gamma} = \OO(l_e,l_{e'}^\epsilon({l_e})).
\eeq
The quantities $l_{e'}^\epsilon({l_e})$ are by construction the Euclidean distances,
in any embedding\footnote{Note that $D\Delta_k$ can always be embedded in $\RR^3$.}
of $D\Delta_k$ in $\RR^3$,  between the vertices of $\Gamma$
that are not connected by the edges of the triangulation.

Thirdly, $\tilde{I}_{\Gamma} \equiv \langle \tilde{O}_{\Gamma} \rangle_{D\Delta_k}$ turns out to coincide with the Feynman amplitude $I_{\Gamma}$ given by
(\ref{Feyamp}). In order to see this, let us express the  gauge fixing terms for the triangulation $D\Delta_k$. We need to choose
a maximal assignment $T_s$ of
$D\Delta_k$ and a assignment $T_l \subset T_s$ maximal in $D\Delta_k \setminus \Gamma$. Notice first that, since all the vertices of $D\Delta_k$ belong to
the graph $\Gamma$, $T_l$ is empty.
Next, recall that, as emphasized in part \ref{inv}, the triangulation
$\Delta_k$ of the ball is dual to a tree $T$; we can take this tree to be
a gauge fixing tree $T_s$, according to the  procedure described in  \ref{sym3d}.
Vertices, edges and simplices of $T_s$ are the vertices, edges and simplices of $\Delta_k\subset D\Delta_k$.
Therefore the gauge fixing is performed by inserting
\beq \label{gfDeltak}
\delta^{T_{s}}_{GF} = \prod_{e\in \Delta_k} (2\pi) \delta_{s_{e},s_{e}^{o}}\prod_{v \in \Delta_k} \chi(\Omega_v^{\epsilon}) \quad ; \quad
D^{T_{s}}_{FP} = \frac{\prod_{\tau \in \Delta_k} \VV_{\tau}}{\prod_{e\in \Delta_k}  l_{e}}
\eeq
Since $\Delta_k$ has no internal vertex or edge, $\Delta_k$ and $D\Delta_k$ possess the same number
of vertices and edges, all located on the boundary of the ball $B$, whereas the number of tetrahedra in $D\Delta_k$ is $2k$. For each tetrahedron $\tau$
in the interior of $B$, there exists a copy of $\tau$ in the exterior of $B$, sharing its edges with $\tau$, and consequently having the same volume.
Moreover, the tetrahedron $\tau$ and its copy have independent orientations $\epsilon_{\tau}$ and $\epsilon'_{\tau}$.
Therefore, using (\ref{gfDeltak}), $\tilde{I}_\Gamma$ can written as a quantity computed on the triangulation $\Delta_k$
\beq \label{Itilde}
\tilde{I}_\Gamma = \int  \prod_{e\in \Delta_{k}} l_e \extd l_e \sum_{\epsilon, \epsilon'\in{\{\pm1\}}^{k}}
\OO(l_e,l_{e'}^{\epsilon,\epsilon'}({l_e}))\left(\prod _{\tau\in \Delta_{k}} \frac{1}{\VV_{\tau}}\right)
 e^{\imath \sum_e s_e^o \omega_e^{\epsilon,\epsilon'}} \prod_{v \in \Delta_k} \chi(\Omega_v^{\epsilon, \epsilon'})
\eeq
where deficit angles are given by
\beq
\omega_e^{\epsilon,\epsilon'}(l_e) = \sum_{ \stackrel{\tau\in \Delta_k}{ \tau \supset e}}(\epsilon_{\tau}+\epsilon'_{\tau}) \theta_e^{\tau}
\eeq
Finally it's easy to convince oneself that the total algebraic solid angles seen from vertices $v$ vanish modulo $4\pi$ only if
$\epsilon_{\tau} + \epsilon'_{\tau} = 0$ for every tetrahedron of $\Delta_k$. Therefore $\tilde{I}_\Gamma = I_{\Gamma}$.

\section{Algebraic structure}\label{alg}

In order to complete our investigation,
we now want to give an algebraic interpretation of the topological state sum model
introduced before. This will show that this model is really a spin foam model, namely constructed
purely in terms of algebraic data.

\subsection{A Poincar\'e model}

\subsubsection{The result}

We first introduce some notations. We define
\be\label{planch}
 \rho \equiv(l_e,s_e),\quad \int \extd\mu(\rho_e) \equiv \frac{1}{2\pi} \sum_{s_e} \int \extd l_e l_e^2
\ee
and introduce a symbol associated to the tetrahedron $\tau$ with vertices $0,1,2,3$:
\be\label{symb}
\left\{\begin{array}{ccc}
\rho_{23}&\rho_{13}&\rho_{12} \\
\rho_{01}&\rho_{02}&\rho_{03}
\end{array}
\right\} \equiv \sum_{\epsilon} \frac{e^{\imath \epsilon_{\tau} S_{\tau}(s_{IJ},l_{IJ})}}{\VV(l_{IJ})}=
2 \frac{\cos S_{\tau}(s_{IJ},l_{IJ})}{\VV(l_{IJ})},
\ee
with \be S_{\tau}(l_{e},s_{e})= \sum_{e\in \tau} s_{e} \theta^\tau_{e}(l_{e}).\ee
The previous model can be written as a Ponzano-Regge like model \cite{Re2}
\be
Z_{\Delta} = \int \prod_{e}\extd\mu(\rho_{e}) \prod_{\tau}\left\{\begin{array}{ccc}
\rho_{e_{1}^\tau}&\rho_{e_{2}^\tau}&\rho_{e_{3}^\tau} \\
\rho_{e_{4}^\tau}&\rho_{e_{5}^\tau}&\rho_{e_{6}^\tau}
\end{array}
\right\},
\ee
where the integral is over all edge labels and where ${e_{i}^\tau}$ label the $6$ edges of $\tau$.

The topological invariance of this model is essentially due to the  identity  (\ref{2,3}), which can rewritten in terms of this symbol as a pentagonal
or Biedenharn-Elliot identity
\begin{equation} \label{pentagon}
\left\{
\begin{array}{ccc} \rho_{1} & \rho_{2} & \rho_{3} \\
                   \rho_{4} & \rho_{5} & \rho_{6}
\end{array} \right\}
\left\{
\begin{array}{ccc} \rho_{1} & \rho_{2} & \rho_{3} \\
                   \rho_{4}' & \rho_{5}' & \rho_{6}'
\end{array} \right\} =
\int \extd \mu(\rho)
\left\{
\begin{array}{ccc} \rho_{5} & \rho_{1} & \rho_{6} \\
                   \rho_{6}' & \rho & \rho_{5}'
\end{array} \right\}
\left\{
\begin{array}{ccc} \rho_{6} & \rho_{2} & \rho_{4} \\
                   \rho_{4}' & \rho & \rho_{6}'
\end{array} \right\}
\left\{
\begin{array}{ccc} \rho_{5} & \rho_{3} & \rho_{4} \\
                   \rho_{4}' & \rho & \rho_{5}'
\end{array} \right\}.
\end{equation}

All these considerations clearly suggest an interpretation of this symbol as a  $6j$-symbol for  a classical group.
This is indeed the case, according to the following

\bigskip

\textbf{Theorem:}
\textit{ $\rho=(l,s)$ labels the unitary representation of the $3$-dimensional (Euclidean) Poincar\'e group $\mathrm{ISO}(3)$
with mass $l$, and spin $s$; $\extd \mu(\rho)$ is the Plancherel measure which arises in the decomposition of $L^2(\mathrm{ISO}(3))$,
and the symbol $\left\{\rho_{i}\right\}$ defined in (\ref{symb}) is the normalized $6j$-symbol of $\mathrm{ISO}(3)$.}

\bigskip

As a first corollary of this theorem, we get an other independent and non geometric proof of the pentagon identity (\ref{2,3}),
since this is a direct consequence of recoupling theory.
An other corollary of this theorem is the fact that this $6j$-symbol satisfies the orthogonality relation
\be \label{ortho}
\int \extd\mu(\rho_{12})
\left\{\begin{array}{ccc}
\rho_{23}&\rho_{13}&\rho_{12} \\
\rho_{01}&\rho_{02}&\rho_{03}
\end{array}\right\}
\left\{\begin{array}{ccc}
\rho_{23}&\rho_{13}&\rho_{12} \\
\rho_{01}&\rho_{02}&\rho_{03}'
\end{array}\right\}
=
2\pi \delta_{s_{03},s_{03}'}\frac{\delta(l_{03}-l_{03}')}{l_{03}^2}.
\ee
It is interesting to give a direct and independent geometric proof of this identity.
We consider two tetrahedra $\tau$, $\tau'$ with vertices $0,1,2,3$,  with edge lengths and spin labels $(l_{IJ},s_{IJ})$,
$(l_{IJ}',s_{IJ}')$, these labels being such that
$l_{IJ}=l_{IJ}',\, s_{IJ}= s_{IJ}'$ except for $(IJ)=(03)$.
We want to compute
\beqa
I &=& \frac{1}{2\pi} \sum_{s_{12}}\int dl_{12} l_{12}^{2} \frac{\sum_{\epsilon} e^{i\epsilon S_{\tau}}}{\VV_{\tau}}
\frac{\sum_{\epsilon'} e^{i\epsilon' S_{\tau'}}}{\VV_{\tau'}}\\
&=& \int dl_{12} l_{12}^{2} \sum_{\epsilon, \epsilon'} \frac{\delta(\epsilon \theta_{12}-\epsilon'\theta_{12}')}{\VV_{\tau}\VV_{\tau'}}
\left(\prod_{(ab)\neq (12)}e^{\imath (\epsilon s_{ab} \theta_{ab}-\epsilon' s_{ab}' \theta_{ab}')}\right)
\eeqa
where $\theta_{ij}$ denotes the dihedral angle at the edge $(ij)$.
One can now use (\ref{dertl})
\be
\frac{\partial \theta_{03}}{\partial l_{12}} = \frac{l_{12}l_{03}}{\VV_{\tau}}, \,\,\, \mathrm{and}\, \,\,
\frac{\partial \theta_{12}}{\partial l_{03}} = \frac{l_{12}l_{03}}{\VV_{\tau}}
\ee
to express the change of integration variables
\be
\extd l_{12} l_{12} = \extd \theta_{03} \frac{\VV_{\tau}}{l_{03}}
\ee
and
\be
\delta(\epsilon \theta_{12}-\epsilon'\theta_{12}')= \delta_{\epsilon,\epsilon'} \delta(l_{03}-l_{03}')\left|\frac{\partial \theta_{12}}{\partial l_{03}}\right|
 = \delta_{\epsilon,\epsilon'}\delta(l_{03}-l_{03}') \frac{\VV_{\tau}}{l_{12}l_{03}}
\ee
Since $l_{03}=l_{03}'$, all the volume factors in the integral cancel out; and since $s_{ab}=s_{ab}'$ for $(ab)\neq (03)$, the integral reduces to
\be
I= \frac{\delta(l_{03}-l_{03}')}{l_{03}^2} \int_{0}^\pi\extd\theta_{03} \sum_{\epsilon} e^{\imath \epsilon \theta_{03}(s_{03}-s_{03}')}
= \frac{\delta(l_{03}-l_{03}')}{l_{03}^2} 2\pi \delta_{s_{03}, s_{03}'}
\ee
which is the desired result.

\subsubsection{Proof of the Theorem}

We will need some definitions. We start by describing the representation spaces of the Poincar\'e group.
Let us fix a frame $(\vec{e}_x,\vec{e}_y,\vec{e}_z)$ of $\RR^3$.
The carrier space corresponding to a representation\footnote{The spin $s \in \frac12\mathbb{Z}$ label representations of the little group $U(1)$.
In the following we restrict to the case where all spins are integers.}
$\rho\equiv (l,s)$ of $ISO(3)=SO(3)\ltimes\mathbb{R}^3$ is given by:
\beq
\mc{E}_{\rho} = \{\phi \in \mc{L}^2(SU(2))\, / \, \forall \xi
\in \left[-\pi,\pi\right], \phi(xh_{\xi}) = e^{i s\xi} \phi(g)\},
\eeq
where $h_{\xi}= e^{\imath \xi\sigma_{3}/2}$ are elements of the $U(1)$-subgroup of rotations with axis $z$ and angle $\xi$.
The action of a rotation $g\in SU(2)$ and a translation $\vec{a}\in \mathbb{R}^4$ on this space is then
\be
(g, 0)\,\phi(x) = \phi(g^{-1}x), \quad
(1,\vec{a})\,\phi(x) =  e^{i\vec{P}_x . \vec{a}}\, \phi(x)
\ee
where $\vec{P}_x\equiv x\cdot\vec{l} $ is the image, by the rotation associated to $x$, of a fixed vector  $\vec{l}= l e_{z}$ of length $l$.

We now define the Clebsh-Gordan coefficients.
The tensor product $(l_{1},s_{1})\otimes(l_{2},s_{2})$ decomposes on representations $\rho_{3}=(l_{3},s_{3})$ such that
$l_{1},l_{2}, l_{3}$ satisfy triangular inequalities and such that the sum of $s_{i}$'s is an integer
\[
 \left|l_{1}-l_{2}\right| \leq l_{3} \leq l_{1} + l_{2}, \quad s_{1}+s_{2}+s_{3} \in \mathbb{Z}
 \]
Given three lengths satisfying these inequalities, one can construct an oriented triangle in the $(x,z)$ plane such that
\be \label{al}
\vec{l}_{3}= a_{\theta_{13}}\cdot \vec{l}_{1} + a_{\theta_{23}}^{-1}\cdot\vec{l}_{2}
\ee
where  one introduces the $2\times2$ matrix representing a rotation of angle $\theta$ and axis $y$
\[
a_{\theta} =
\left( \begin{array}{cc}
\cos \theta/2&\sin \theta/2 \\
-\sin \theta/2&\cos \theta/2
\end{array} \right). \]
The angles $\theta_{13}, \theta_{23} \in \left[0,\pi\right]$ are the angles between
the edges $(1)$ and $(3)$ (resp. $(2), (3)$) within the oriented triangle $\left[123\right]$ whose edges lengths are $l_1, l_2, l_3$.
They are characterized by
\be
l_{1}^2 = l_{2}^2 + l_{3}^2 -2l_{2}l_{3} \cos\theta_{23}
\ee
and similarly for $\theta_{13}$.
The Clebsh-Gordan coefficient:

\begin{figure}[h]
\begin{center}
\psfrag{V}{$\equiv \C{i}{j}{k}$}
\includegraphics[width=3cm]{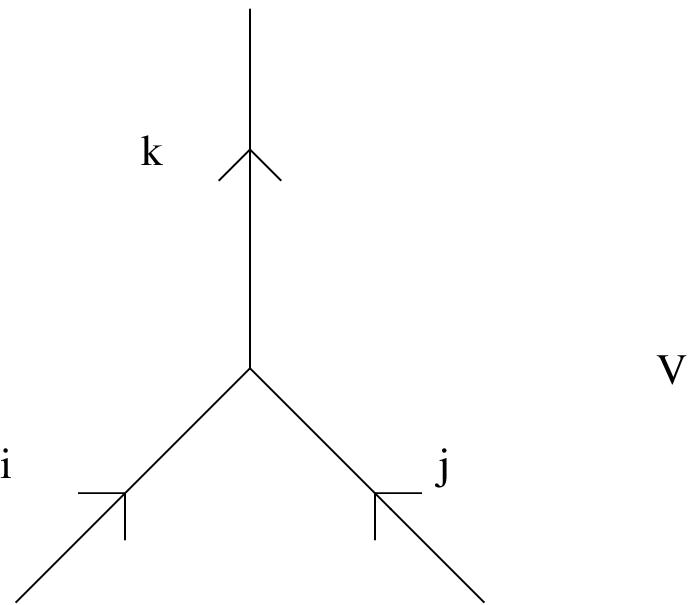}
\end{center}
\end{figure}
\noindent also called $3j$-symbol, is defined (up to a normalization factor) to be the invariant map
$\mc{E}_{l_{i},s_{i}}\otimes\mc{E}_{l_{j},s_{j}} \rightarrow \mc{E}_{l_{k},s_{k}}$.
The structure of this object is similar to the Clebsh-Gordan coefficient of the quantum double $\mathcal{D}SU(2)$,
whose construction is described in \cite{bm}.
When properly normalized, it can be written as
\be
\label{3j}
\C{1}{2}{3}
= \left(\frac{l_1l_2l_3}{\pi}\right)^{-\frac12} \int_{U(1)} \prod_{i=1}^3(\extd{\zeta}_i e^{\imath s_i \zeta_i})\,
\delta_{x_1h_{\zeta_1}^{-1}}(x_3h_{\zeta_3}a_{\theta_{13}})\delta_{x_2h_{\zeta_2}^{-1}}(x_3h_{\zeta_3}a_{\theta_{23}}^{-1})
\ee
where $\extd{\zeta}$ is the normalized Haar measure on $U(1)$.
It is clear that this coefficient is such that $C^{\rho_{1}\rho_{2}\rho_{3}}_{x_{1}h_{\xi_{1}}x_{2}h_{\xi_{2}}x_{3}h_{\xi_{3}}}
= \C{1}{2}{3} e^{-\imath(s_1 \zeta_1 + s_2\zeta_2 -  s_3\zeta_3)}$.
Moreover, it is straightforward to check (thanks to (\ref{al})) that it
defines an invariant tensor, that is, $C^{\rho_{i}}_{gx_{i}}= C^{\rho_{i}}_{x_{i}}$,
and that $\sum_{i}x_{i}\cdot \vec{l_{i}}$=0.
The condition that imposes the sum of the spins to be an integer comes from the identities
$C^{\rho_{i}}_{x_{i}}=C^{\rho_{i}}_{-x_{i}}=C^{\rho_{i}}_{x_{i}h_{2\pi}}= e^{2i\pi \sum_{i}s_{i}} C^{\rho_{i}}_{x_{i}}$.

The normalization factor is chosen in order to have the resolution of the identity \cite{PR2}:
\beq
 \int \extd\mu(\rho_{3})  \left(\int\extd x_3 \C{1}{2}{3} \overline{C_{z_1z_2x_3}^{\rho_{1}\rho_{2}\rho_{3}}}\right)
= \bar{\delta}_{z_1}^{s_1}(x_1)\bar{\delta}_{z_2}^{s_2}(x_2)
\eeq
where the `bar' denotes the complex conjugate, $\extd\mu(\rho_{3})$ denotes the Plancherel measure (\ref{planch}), and
\[
\bar{\delta}_{z_i}^{s_i}(x_i) = \int_{U(1)} \extd \zeta e^{-\imath s_i \zeta} \delta_{z_ih_{\zeta}}(x_i)
\]

It is worth mentioning two further properties that the Clebsh-Gordan (\ref{3j})
satisfies. The first one is a `braiding' property  which describe the change of the Clebsh-Gordan under the interchange of
the labels $i$ and $j$:
\beq
\label{braid}
\C{2}{1}{3} = (-1)^{s_1+s_2-s_3} \C{1}{2}{3},
\eeq
derived thanks to the identity $h_{\pi}a_{\theta}h_{-\pi}= a_{-\theta}$.
The second property expresses the dual intertwiner $\tilde{C}^{\rho_{1}\rho_{2}\rho_3}: \mc{E}_{\rho_3} \rightarrow \mc{E}_{\rho_{1}}\otimes \mc{E}_{\rho_{2}}$
in terms of the $3j$-symbol (\ref{3j}) :
\beq
\label{dual}
\tilde{C}^{\rho_{i}\rho_{j}\rho_{k}} = \overline{C^{\rho_{i}\rho_{j}\rho_{k}}}.
\eeq

Finally, it will be important to note that, if one restricts, as we do, the spin to be integers, one can restrict the integral in (\ref{3j}) to be over
$\xi\in [0, 2\pi]$, provided that one takes the $SO(3)$
delta function in (\ref{3j}), namely $\delta_{g}(x)\equiv \frac12(\delta_{g}(x)+\delta_{-g}(x))$.

One can now construct the $6j$-symbol, defined as a contraction of Clebsh-Gordan coefficients
\be
\label{6jsymb}
\int \extd x_{01} \extd x_{02}\extd x_{03}\, \overline{\C{12}{02}{01}} \C{03}{13}{01} \C{02}{23}{03}
\equiv \left\{\begin{array}{ccc}
\rho_{23}&\rho_{13}&\rho_{12} \\
\rho_{01}&\rho_{02}&\rho_{03}
\end{array}
\right\} \C{23}{13}{12}.
\ee
The way the contractions occur is dictated by the geometry of a tetrahedron $(0,1,2,3)$ with oriented edges $(IJ)$. There is one Clebsh-Gordan
attached to each face of the tetrahedron, and one representation for each edge.


\medskip

Let us now prove the theorem, by first noting that
each Clebsh-Gordan of the LHS of (\ref{6jsymb}) supplies two $\delta$-functions on $SU(2)$, so the integral contains $6$ delta functions.
The integration is over three $SU(2)$ elements $x_{0i}$, and
$6$ $U(1)$ elements, denoted by $\xi_{0i}, \xi_{0i}'$. By performing a simple change of variables $x _{0i}\to x _{0i}h_{\xi_{0i}}$,
one sees that the integrand depends only on the $x_{0i}$ and the differences $\xi_{0i} - \xi_{0i}'$.
Then the integration over the $SU(2)$ variables allows us to solve straightforwardly three of the delta functions.
In order to write the result of integration in a convenient form,  we assign an angle $\xi_{IJ}$ to each edge $(IJ)$, $0\leq I, J\leq 3$ of the tetrahedra
$\left[0123\right]$,
and denote the corresponding group element by $h_{IJ} = h_{JI} \equiv h_{\xi_{IJ}} $.
We also denote by $a_I^{\,JK}=a_I^{\,KJ}$  the rotation $a_{\theta_{(IJ)(IK)}}$ associated to the geometric angle
between the edges $(IJ)$ and $(IK)$ in the triangle $\left[IJK\right]$. The later angles are fully determined by the lengths $l_{IJ}$.
After integration,  the LHS of (\ref{6jsymb}) takes the form
\be\label{ev6}
\int \prod_{I<J}\left(\extd \xi_{IJ} e^{-\imath s_{IJ} \xi_{IJ}} \right)
\,\delta(h_{01}a_0^{\,12}h_{02}a_0^{\,23}h_{03}a_0^{\,31})
\delta_{x_{13}h_{13}^{-1}}(x_{12} h_{12} g_{1})
\delta_{x_{23}h_{23}^{-1}}(x_{12} h_{12} g_{2})
\ee
times the prefactor $(\pi^{-1}l_{01}l_{02}l_{03})^{-1}(\pi^{-1}l_{12}l_{23}l_{31})^{-1/2}$,
where we have defined
\be
g_{1}\equiv (a_1^{\,02})^{-1}h_{10}a_0^{\,13}h_{03}(a_3^{\,02})^{-1}, \quad
g_{2}\equiv  (a_1^{\,20})^{-1}h_{10}(a_1^{\,03})^{-1}.
\ee
One can simplify the expression for $g_{1}$ if one remembers that the sum of the three interior angles of a flat triangle is equal to $\pi$.
Using this fact for the triangles $\left[012\right]$ and $\left[023\right]$, we get
$(a_1^{\,02})^{-1} = a_{-\pi}a_2^{\,01}a_0^{\,21}$ and $(a_3^{02})^{-1}= a_{-\pi} a_0^{\,32}a_2^{\,03}$. Plugging this into the expression for $g_{1}$
yields
\be
g_{1}= a_{-\pi} a_{2}^{\,10}\left\{ a_0^{\,12}h_{01}a_0^{\,13}h_{03}a_0^{\,32} h_{02} \right\}h_{20}^{-1}a_2^{\,03}a_{-\pi}
= a_{-\pi} a_{2}^{\,10}h_{20}^{-1}a_2^{\,03}a_{-\pi}
=- a_{2}^{\,10}h_{20}a_2^{\,03},
\ee
where for the last two equalities, we first have used that the term inside the bracket is $1$ - this is due to the first delta function in (\ref{ev6})-,
and then the  property $a_{\pi}ha_{\pi}=-h^{-1}$.

In order to complete the proof of the theorem, we first have to find the solutions of the constraint expressed by the
first $\delta$-function in (\ref{ev6}).
These solutions of the constraint arise from the flatness of the tetrahedron, which leads to a remarkable identity at each vertex (see (\ref{fid})):
\beq \label{i}
h^\epsilon_{IJ}  a_I^{\,LI} h^\epsilon_{IL} a_i^{\,LK}h^\epsilon_{IK}a_I^{\,KJ}= -\epsilon \mathbf{1}\quad \forall I
\eeq
$(IJKL)$ is an even permutation of $(0123)$, $h^+_{IJ} \equiv h_{\phi_{IJ}^e}$ is the rotation associated to the exterior dihedral angle of the
edge $(IJ)$.  Furthermore, if one considers (\ref{i}) as an
equation in $\xi_{IJ}$, then $\xi_{IJ}=\phi^{e}_{IJ}$, and this is  the unique solution belonging to the interval
$\left[0,2\pi\right]$.
Multiplying (\ref{i}) by $a_{\pi}$ on the left, and $a_{-\pi}$ on the right, has the effect to change all $h$ into $h^{-1}$, while leaving
the $a$'s unmodified; is one now multiplies the LHS by $h_{2\pi}^3=-1$, one sees that $\xi_{IJ}= 2\pi-\phi^{e}_{IJ}$ is a solution of (\ref{i})
with $+1$ in the RHS.
Theses identities are  carefully studied in appendix \ref{6j}.
The bottom line is that the equation $h_{01}a_0^{\,12}h_{02}a_0^{\,23}h_{03}a_0^{\,31}=\pm 1$ (recall that we work with integer spin) admits
two solutions when $\xi_{IJ}\in [0, 2\pi]$, namely
\be\label{solphi}
\xi_{IJ}^+= \phi^{e}_{IJ}, \,\, \mathrm{or}\,\,\, \xi_{IJ}^-= 2\pi-\phi^{e}_{IJ},
\ee
We denote the respective $h_{\xi_{IJ}^\pm}$ by $h^{\pm}_{IJ}$.

For each of these solutions, we use (\ref{i}) to express the Euler decomposition of $g_{1}$ and $g_{2}$
\beqa
g_{1} &=& (a_1^{\,20})^{-1}h^\epsilon_{10}(a_1^{\,03})^{-1}= -\epsilon ( h^\epsilon_{12})^{-1} a_{1}^{\,23}(h^\epsilon_{13})^{-{1}} \\
g_{2}& =& - a_{2}^{\,10}h^\epsilon_{20}a_2^{\,03}= \epsilon (h^\epsilon_{21})^{-1}(a_{2}^{\,31})^{-1} (h^\epsilon_{23})^{-1}
\eeqa
Using this decomposition one can rescale $\xi_{ij}\to \xi_{ij}-\xi^\epsilon_{ij}$, $1\leq i,j\leq 3$ in (\ref{ev6}), and factorize from it the
Clebsch-gordan coefficient $\C{23}{13}{12}$. The proportionality coefficient for each set of solutions is
\be\label{lastev}
\prod_{i<j}e^{-\epsilon \imath s_{ij}\phi_{ij}^{e}} \left(\frac{\pi}{l_{01}l_{02}l_{03}}\right)
\int_{\epsilon}\left(\prod_{i}\extd \xi_{0i}e^{-\imath \xi_{0i}}\right) \delta(h_{01}a_0^{\,12}h_{02}a_0^{\,23}h_{03}a_0^{\,31})
\ee
where the subscript $\epsilon$ in the integral reminds that we need to integrate around the solution $\xi_{0i}=\xi_{0i}^\epsilon$.

Now if we denote $g(\xi_{0i})\equiv h_{01}a_0^{\,12}h_{02}a_0^{\,23}h_{03}$,
(\ref{measure}) gives the Jacobian of transformation from the Haar measure to the $\xi_{0i}$ measure
around a solution $g(\xi_{0i})a_0^{\,31}= \pm 1$:
\beq\label{Jac}
 \extd g = \pi \frac{\VV(l_{IJ})}{l_{01}l_{02}l_{03}} \prod_i \extd \xi_{0i}
\eeq
where the volume of the tetrahedron appears.
Therefore the expression in (\ref{lastev}) can be written as
\be
\prod_{i<j}e^{-\epsilon \imath s_{ij}\phi_{ij}^{e}} \left(\frac{\pi}{l_{01}l_{02}l_{03}}\right)
   \int\left(\prod_{i}\extd \xi_{0i}e^{-\imath \xi_{0i}}\right)
   \delta(\xi_{0i}-\xi^{\epsilon}_{0i}) \frac{l_{01}l_{02}l_{03}}{ \pi \VV(l_{IJ})}
=\frac{\prod_{I<J} e^{-\epsilon \imath \phi^{e}_{IJ}}}{\VV(l_{IJ})}.
\ee
The $6j $-symbol is eventually obtained by summing the two contributions $\epsilon =\pm$. Therefore it reads
\beq
\left\{\begin{array}{ccc}
\rho_{23}&\rho_{13}&\rho_{12} \\
\rho_{01}&\rho_{02}&\rho_{03}
\end{array}
\right\} = (-1)^{\sum s_{IJ}} 2\frac{\cos S_{\tau}(s_{IJ},l_{IJ})}{\mathcal{V}(l_{IJ})}
\eeq
The exterior dihedral angles have been replaced by the interior ones $\theta^i = \pi - \phi^e$, leading to a factor $(-1)^{\sum s_{IJ}}$ in the formula.

\subsection{Poincar\'e versus Ponzano-Regge}

We have shown in the previous sections that a Feynman graph amplitude is the evaluation of the observable $\OO(l_e)$ for a Poincar\'e spin-foam model.
On the other hand, we know \cite{PR3} that a usual Feynman graph is also the abelian limit of the evaluation of the same observable for the Ponzano-Regge
model. This means that one should be able to relate in a direct way our Poincar\'e model with a suitable semi-classical limit of the Ponzano-Regge model.

In order to do so, let $(u_e, v_e)$ be an edge-labelling of a triangulation $\Delta$, with pairs of half-integers. We consider the square of the
Ponzano-Regge model:
\beq
\label{pr2}
Z^{PR^2}_{\Delta} =  \sum_{\{u_e,v_e\}} \prod_e \extd_{u_e} \extd_{v_e} \prod_{\tau} \{u\}_{\tau} \{v\}_{\tau}
\eeq
where $\{u\}_{\tau}$ and  $\{v\}_{\tau}$ denotes $6j$-symbols of $SU(2)$ and $d_j = 2j+1$. After a simple change of variables
\[ u = j+s \nonumber, \quad v = j-s \]
the partition function reads
\beq \label{prmodif}
Z^{PR^2}_{\Delta} = \sum_{\stackrel{\{j_e,s_e\}}{ |s_{e}|\leq j_{e}}} \prod_e (\extd^2_{j_e} - (2s_{e})^2) \prod_{\tau} \{j+s\}_{\tau} \{j-s\}_{\tau}
\eeq
Next, let us introduce a length variable $l = d_jl_p $, where $l_p$ is the Planck length. We would like to evaluate the limit $l_p \to 0$ of the model, in
a sector where the typical value of $l$ is kept fixed, and the spins $s$ are bounded by a cut-off $n$ much smaller than the typical value of $j$:
$n \ll l/l_p$.

Interestingly, using the techniques presented in \cite{LDA1}, asymptotics of $6j$-symbols in
(\ref{prmodif}) can easily be evaluated in this limit. We get:
\beq
\{\frac{l}{2l_p}  \pm s -\frac12\}_{\tau}^2 \sim  \frac{4l_p^3}{\pi} \frac{1}{\VV(l_{ij})} \cos^2 \left[\sum_{i<j}
\left(\frac{l_{ij}}{2l_p} \pm s_{ij} \right) \theta^{\tau}_{ij} +\frac{\pi}{4}\right]
\eeq
for fixed values of the length $l$ and the spin $s$. This suggests the following semi-classical limit for the model:
\beq \label{semcl}
\frac{1}{2^{|e|}}\left(\frac{l_p^{3}}{\pi}\right)^{(|e|-|\tau|)} Z^{PR^2}_{\Delta} \sim   \frac{1}{(2\pi)^{|e|}}
\int \prod_{e\in\Delta} \extd l_e  \prod_e  l_e^2
\sum_{\{s_e \leq n\}} \sum_{\{\epsilon_{\tau},\epsilon'_{\tau}\}} \prod_{\tau} \frac{e^{i\mathcal{S}_{\tau}(l_p)}}{\mathcal{V}_{\tau}}
\eeq
where $\mathcal{V}_{\tau}$ is the volume of the tetrahedron $\tau$ with edge lengths $l_e$, and the action is given by (we write $\theta_e$ for  $\theta_e^{\tau}$
and $\epsilon,\epsilon'$ for $\epsilon_{\tau},\epsilon^{'}_{\tau}$):
\beq
\mathcal{S}_{\tau}(l_p) = \sum_{e\in\tau}\, \frac{1}{2l_p} (\epsilon+ \epsilon')l_e\theta_e(l_e) + (\epsilon-\epsilon') s_e\theta_e(l_e) + (\epsilon+\epsilon')\frac{\pi}{4}
\eeq
We can reorganize the sum over the orientations\footnote{using $2\cos\frac{a+b}{2}\cos\frac{a-b}{2}= \cos a + \cos b$.}:
\beq
\sum_{\{\epsilon_{\tau},\epsilon'_{\tau}\}} e^{i\mathcal{S}_{\tau}} =
\sum_{\eta_{\tau}=\pm 1}  e^{{i}\eta_{\tau} \left( \sum_{e\in\tau}\frac{l_e}{l_{p}}\theta_e  +
\frac{\pi}{2}\right)} + \sum_{\eta'_{\tau}=\pm 1} e^{i \eta'_{\tau} \sum_{e\in\tau} (2s_e)\theta_e}
\eeq
and identify both Regge action and Poincar\'e action. Denoting the RHS of (\ref{semcl}) by $\tilde{Z}_{\Delta}(n,l_p)$, we are interested in the quantity
\beq
Z_{\Delta} = \lim_{n\to\infty} \lim_{l_p \to 0} \tilde{Z}_{\Delta}(n,l_p)
\eeq
We will show that $Z_{\Delta}$ is the partition function of our Poincar\'e model (\ref{3dmod}). Let us focus on the first limit $l_p \to 0$,
$n$ being fixed. The integrand of $\tilde{Z}_{\Delta}(n,l_p)$ is a product of sums which one can develop.
The integral splits then into $2^{|\tau|}$ terms of
the form
\beq \label{terms}
\frac{\imath^{|K_2|}}{(2\pi)^{|e|}} \int\prod_{e\in\Delta}\mathrm{d}l_el^2_e\sum_{\{s_e \leq 2n\}} \prod_{\tau}\,
\frac{1}{\mathcal{V}_{\tau}}\,\prod_{\tau\in K_1}
\left(\sum_{\eta'_{\tau}} e^{\imath \eta'_{\tau}\sum_{e \in K_1} s_e \theta_e}\right)
\prod_{\tau\in K_2} \left(\sum_{\eta_{\tau}} \eta_{\tau} e^{\frac{\imath}{l_p}\eta_{\tau}\sum_{e \in K_2}l_e \theta_e}\right)
\eeq
$K_1$ and $K_2$ are disjoint sets covering the all set of tetrahedra; $|K_2|$ denotes the cardinal of $K_2$. Note that we have made the change
of variables
$s_e \to 2s_e$, and then the spin are now integers. All terms such that $K_2 \not= \emptyset$ are oscillatory integrals, and therefore under
stationary phase evaluation
 are typically of  order $l_p^{|K_2|/2}$. The term corresponding to $K_2 = \emptyset$
is independent of $l_p$ and is therefore the only term surviving when $l_p$ goes to zero.
\beq
\lim_{l_p \to 0} \tilde{Z}_{\Delta}(n,l_p) = \frac{1}{(2\pi)^{|e|}} \int\prod_{e\in\Delta}\mathrm{d}l_el^2_e\sum_{\{s_e \leq 2n\}} \sum_{\{\eta_{\tau}\}}
\left(\prod_{\tau}\,\frac{e^{\imath \eta_{\tau}\sum_e s_e \theta_e} }{\mathcal{V}_{\tau}}\right)
\eeq
Eventually, by taking the limit $n \to \infty$, we get the desired result and recover the Poincar\'e model
in this semiclassical limit.

\section{Feynman diagrams on homogeneous spaces}

In this section we consider a scalar quantum field theory on spherical space-time.
A Feynman amplitude of a graph $\Gamma$ embedded in
the unit $3$-sphere $\mathcal{S}^3$ takes the form
\beq \label{Feysph}
I_{\Gamma} = \int_{S^3} \extd u_1 \cdots \extd u_N \prod_{(ij) \in \Gamma} G_m (l_{ij})
\eeq
$\extd u_i$ is the normalized measure on the $3$-sphere. The integrand is a product of propagators, which are functions of the dimensionless spherical
distances $l_{ij} \in \left[0,\pi\right]$ between the vertices, and
invariant under the action of the group $SO(4)$.

The results obtained for the flat case can directly be extended to the spherical one.
Again, the idea is
to gauge out the integration measure. The invariant measure is expressed in terms of spherical lengths and the Haar measure $dU$ of $SO(4)$
in the same way as in section \ref{inv}.
\beq
\extd u_1 \cdots \extd u_{3+k} = \extd U \sum_{\epsilon \in \{\pm 1\}^k} \prod_{e\in \Delta_{k}} \extd l_e \sin l_e
   \prod_{\tau\in\Delta_{k}} \frac{1}{\mathcal{V}_{\tau}}
\eeq
$\Delta_k$ is a triangulation, with spherical tetrahedra, of a $3$-ball in $\mathcal{S}^3$, which possesses $3+k$ vertices on its boundary, and
which does not contain any internal edge. $\mathcal{V}_{\tau}$ is the square root of the Gram determinant $\det\left[\cos\,l_{ij}\right]$ (see appendix)
associated to the tetrahedron $\tau$.

A key identity relates the Gram determinants of the tetrahedra of a spherical $4$-simplex, which has a form analogous to (\ref{id3d}). We refer the reader
to the appendix where the formula (\ref{lemm1}) and (\ref{lemm2}) are established for spherical simplices. This leads to the emergence
of the topological spin-foam model
\beq \label{modsph}
Z_{\Delta} = \frac{1}{(2\pi)^{|e|}}\int\prod_{e\in\Delta} \mathrm{d}l_e \sin^2 l_e \sum_{\{s_e,\epsilon_{\tau}\}}\left(\prod_{\tau}
\frac{e^{\imath \epsilon_{\tau} S_{\tau}(s_{e},l_{e})}}{\mathcal{V}_{\tau}}\right)
\eeq
The action for each tetrahedron reads
\beq \label{actsph}
S_{\tau} = \sum_{e\in \tau} s_e \theta_e^{\tau}(l_e)
\eeq
where $\theta_e^{\tau}$ is the spherical interior dihedral angle of the edge $e$ in $\tau$.
The Feynman amplitude $I_{\Gamma}$ is then the expectation value of the observable
$\OO(l_e) = \prod_{e \in \Gamma} G_m(l_e)$ for the model (\ref{modsph}) computed on any triangulation $\Delta$ of $\mathcal{S}^3$
which contains $\Gamma$ as a subgraph. Analogous results for hyperbolical space are obtained by working on the hyperboloid and by
replacing all the angles by hyperbolic angles.

Again, the model (\ref{modsph}) takes the form of a
Ponzano-Regge model. The weight associated to this model satisfies a pentagonal identity, and can be interpreted as a $6j$-symbol.
It is not difficult to convince oneself that the underlying group is the quantum double $\mathcal{D}SU(2)$ \cite{bm};
the proof of this statement is identical to the flat case, so we do not repeat it here.

The classical study of the action (\ref{actsph}) differs from the flat case, since the eom non longer admit
the so called Regge solutions $s_e=\alpha l_e$, unless $\alpha = 0$. Therefore the connection with usual Regge gravity seems to be lost.
However, it is worth noting that these solutions can be reintroduced by
adding a volume term $2\sum_{\tau} \mbox{Vol}_{\tau}(l_e)$ to the action (\ref{actsph}), where
$ \mbox{Vol}_{\tau}(l_e)$ is the spherical volume of the tetrahedra (which should not be confused with ${\mathcal{V}_{\tau}}$).

One can check that the model obtained is still topological, and that Feynman amplitudes can be written as evaluations of observables
for this model as well.
Let us introduce length variables $l_e = \sqrt{\Lambda} L_e$, where $\Lambda$ is the cosmological constant. The new action
can be written in terms of the geometrical data of a
tetrahedron of curvature $\Lambda$
\beq \label{act}
S_{\tau_{\lambda}} = \sum_{e\in \tau_{\Lambda}} s_e \theta_e^{\tau_{\Lambda}}(L_e) + 2 \Lambda^{3/2} \mbox{Vol}_{\tau_{\Lambda}}(L_e)
\eeq
where $\tau_{\Lambda}$ is the image of $\tau$ by the re-scaling $l_e \rightarrow L_e=l_e/\sqrt{\Lambda}$, and $\theta_e^{\tau_{\Lambda}}$
is the dihedral angle of $e$ in
$\tau_{\Lambda}$. Now using the spherical Schl\"afli identity
for the tetrahedron $\tau_{\Lambda}$ $$ \sum_{e} L_e \extd \theta_e + 2 \Lambda \extd \mbox{Vol}= 0,$$
we see that the variation of the action (\ref{act}) gives Regge solution $s_e = \sqrt{\Lambda} L_e$.
If we insert this solution
into the action (\ref{act}), we recover the Regge action describing discrete 3d gravity with cosmological constant $\Lambda$.
Thus, the presence of a cosmological constant removes the degeneracy of the Regge solutions, and leads to the natural value $\alpha = \sqrt{\Lambda}$
for the
scale parameter. Eventually let us remark that, since the topological model associated to the action (\ref{act}) reduces to the Poincar\'e model
(\ref{id3d}) when $\Lambda \to 0$, we expect the associated symbol to
be, again, the $6j$-symbol of a deformation of the Poincar\'e group.

\section{Gravity and BF theory}

Before concluding, we would like to give some continuum explanation for the results we have obtained here.
The fact that our state-sum model is built from $6j$-symbols of the Poincar\'e group shows, from standard arguments \cite{rev}, that the topological theory
which we have described in terms of a spin foam model is the quantization of a Poincar\'e $BF$ theory.
This theory can be written in terms of a one-form $E$ and a connection $A$, both valued in the Poincar\'e algebra,
whose generators $J_{i}, P_{i}$, $i=0,1,2$ satisfy
\be
           [J^{i},J^{j}] = 2\epsilon ^{ijk}J_{k}, \quad [J^{i},P^{j}] = 2\epsilon ^{ijk}P_{k}, \quad [P^{i}, P^{j}] =0,
\ee
 and
\be
E \equiv e^{i}P_{i} + b^{i}J_{i}, \quad A\equiv c^{i}P_{i} + \omega^{i}J_{i}.
\ee
The action is given by
        \be
            S=\int_{M}tr({E}\wedge F[{A}])
        \ee
where $F(A)=dA + A\wedge A$, and where the invariant trace is taken to be $\mathrm{tr}(J_{i}P_{j})= \delta_{ij}$,
$\mathrm{tr}(P_{i}P_{j})=\mathrm{tr}(J_{i}J_{j})=0$.
In components this reads
\be
S= \int_{M}(e_{i}\wedge R^{i}[\omega]+b_{i}\wedge d_{\omega}c^{i})\label{sbcea}.
\ee
This model was introduced in \cite{carlip} and recently studied in \cite{MP}.
$e$ is interpreted as the three dimensional frame field, $\omega$ as the spin connection;
$b,c$ are some topological matter fields.
Its equations of motion when $b=c=0$ have, as the only solution, the flat space-time
$d_{\omega}e=0$ and $R(\omega)=0$.
Following \cite{PR1}, one also expects the Feynman diagram observables to be related to Wilson lines observables for this theory
\be
\OO = \mbox{P} \mbox{exp} \int_{\Gamma}\mathrm{tr}(e_{t}J_{0}).
\ee

This result is reminiscent to the Polyakov reformulation of Feynman amplitudes as worldline integrals \cite{Polyakov}. Such an interpretation provided a natural way, by replacing worldlines by worldsheets integrals, to deform field theory in terms of a dimensionfull parameter - leading to string theory.
Analogously, our formulation of Feynman amplitudes in terms of spin foam models may provide a new way to think about consistent dimensionfull deformation of field theory structure, in order to take into account the quantum nature of background geometry.

\section{Conclusion}

We have given in this paper a new perspective on closed Feynman diagrams in three dimensions.
The main point was to show that the language of spin foam models - which was developed in order to address the problem of a background independent
approach to
quantum  gravity - naturally arises in the context of quantum field theory, and that a careful study of Feynman amplitudes can even lead to a
purely algebraic understanding of
the quantum weights. It was also shown that usual field theory can be given a background independent perspective in the sense that the flat
space geometry is purely encoded
in terms of a choice of  weights controlling the dynamics of the geometry .

This result is by itself very encouraging since it clearly shows that the language of spin foam is not unrelated
to the usual language of field theory and also suggests a new way to investigate the semi-classical limit of quantum gravity.
Namely, the Feynman diagram spin foam model provides a `zero order' approximation of the full quantum gravity model, giving
strong constraints on the possible quantum gravity extensions. Such consistent extensions are known in 3d; but it should be clear from the analysis presented here and as shown in \cite{4d} that the results of this paper can be extended to higher dimensions.

We have addressed in this paper the case of scalar field closed Feynman diagrams. This clearly needs to be investigated further;
first to the case of open
diagrams but also to the case of higher spin fields. We expect in the later extension, the variable $s$ which appears in our model, to acquire a
more direct physical meaning.
It would also be useful  to have a deeper understanding of how usual Feynamn diagrams can be understood as the expectation value of topological
observables from the point of view of the continuum,
as sketched in the last section.

\bigskip

{\bf Acknowledgments:}
We are grateful to the organizers of the conference Loops '05, Potsdam October 2005, where A. B. presented part of these results.
We would like to thank R. Williams, T. Konopka, E. Livine and L. Smolin for useful discussions.
A. B. would like to thank the Perimeter Institute for hospitality.
This work is partially supported by a MENRT graduate student grant, by Eurodoc program from R\'egion Rh\^one-Aples and by
the Government of Canada Award Program.

\appendix

\section{Geometry of simplicial complexes}  \label{geo}

In this appendix we first recall some definitions and notations, and give a proof of the results (\ref{lemm1}), (\ref{lemm2}).

\subsection{Basic definitions and discussion about  orientation}

We start by recalling basic features about $D$-dimensional simplices. An abstract (or combinatorial) $D$-simplex is a $D+1$-elements set
$\{0,\cdots, D\}$. Its $k+1$-elements subsets  are called $k$-faces of the simplex.
An embedded spherical $D$-simplex is defined by $D+1$  normalized vectors  $(e_0,\cdots e_D)$ in $\RR^{D+1}$, $e_{i}^2= 1$.
The spherical edge lengths $l_{ij}$ are given by \[ \cos l_{ij}= e_{i}\cdot e_{j}.\]
We denote by $(b_0,\cdots,b_D)$ the dual vectors: $b_i.e_j\,\equiv \,\delta_{ij}$, and we define :
\beqa
\mathcal{V}(e_0,\cdots,e_D)\,&\equiv&\,|\mbox{det}(e_0,\cdots,e_D)|\\
G(e_0,\cdots e_D)\,&\equiv&\,\mbox{det}(e_i.e_j)\,=\mbox{det}(\cos l_{ij})=\,\mathcal{V}^2(e_0,\cdots,e_D)
\eeqa
$\mathcal{V}^2 = G$  is the determinant of the Gram matrix associated to this simplex.
The  \textit{interior} dihedral angles are denoted by $\theta_{ij}\in [0,\pi]$ and defined by
\[
b_i.b_j\,=\,-\left|b_i\right|\left|b_j\right|\,\mbox{cos}\, \theta_{ij}
\]
In this paper we work only with the interior angles and refer to them simply as dihedral angles. $\theta_{ij}$ is the
angle between the $(D-1)$-faces $i$ and $j$ (i.e the faces obtained by dropping the points $i$ or $j$)
\footnote{The exterior dihedral angle given by $\phi_{ij}\equiv \pi-\theta_{{ij}}$ is the angle between the normals to the $(D-1)$-simplices $i,j$.}.
Hence, they are labelled by the $(D-2)$-face $F_{ij}$ opposite to $i$ and $j$. In particular for
$D=3$, dihedral angles are labelled by the edges of the tetrahedron, and for $D=4$ by the faces of the $4$-simplex.

\medskip

An orientation of an abstract $D$-simplex is a choice of an ordering $(0,\cdots D)$ of its
points, up to even permutations \cite{topo}. A coloring $(l_e, \epsilon)$ of an oriented simplex is a set of numbers $l_e \in \left[0,\pi\right]$,
labelling the edges of the simplex and
satisfying triangular inequalities, together with a sign $\epsilon = \pm1$. A given coloring promotes an abstract oriented simplex to an embedded spherical
simplex; namely, it defines, up to rotations in $SO(D+1)$, normalized vectors $(e_0,\cdots e_D)$ - ordered according to the orientation -  in $\RR^{D+1}$,
such that $\cos l_{ij}= e_{i}\cdot e_{j}$, and such that the sign of the determinant $\mbox{det}(e_0,\cdots e_D)$ is $\epsilon$.
The embedding is then called \textit{compatible} with the coloring.
In the flat case, a coloring $(l_e, \epsilon)$, with $l_e \in \mathbb{R}_+$ and $\epsilon = \pm1$,
specifies the geometry of a flat simplex $(0,\cdots D)$ in $\RR^D$, namely a set of $D$ vectors $\vec{l}_i, \,i=1 \cdots D$ represented
by the oriented edges $(0i)$, and such that the sign of the determinant $\mbox{det}(\vec{l}_1,\cdots \vec{l}_D)$ is $\epsilon$.
For instance, $4$ ordered points $(0,1,2,3)$, six
positive numbers $l_{ij}, \, i,j = 0 \cdots 3$ and an element $\epsilon \in \{\pm1\}$, determine the geometry of a flat tetrahedron in $\RR^3$.

An orientation of a $D$-simplex induces an orientation for each of its $(D-1)$-faces by dropping the
missing point in an even ordering of the points of the $D$-simplex in which this missing point is first. Now lets consider two oriented
$D$-simplices sharing a $(D-1)$-face. The orientations of these $D$-simplices are said consistent with each other if the shared face
inherits opposite orientations from them.
A triangulation $\Delta$ is said orientable if there exists a choice of consistent orientations for all simplices \cite{topo};
such a choice is then, by definition, an orientation of $\Delta$.
In this work we consider only orientable triangulations.
For each of the triangulations we are working with, an orientation is implicitly chosen and fixed; and we don't refer to it anymore.
A coloring of a triangulation is then a set
$\{l_e, \epsilon_{\sigma}\}$, where $l_e \in \left[0,\pi\right]$ label the edges and $\epsilon_{\sigma} \in \{\pm1\}$ label the simplices.
Abusing terminology, the signs $\epsilon_{\sigma}$ will be called `orientations' of the simplices.

With these definitions it is now straightforward to check the property illustrated in Fig \ref{convD2},
section \ref{inv}. Given two $D$-simplices  $\sigma_1$ and $\sigma_2$ glued along a $(D-1)$-face, a coloring
$c_{\sigma_1\cup\sigma_2} = \{l_e, \epsilon_{\sigma_1}, \epsilon_{\sigma_2}\}$ defines a map
$\phi_{\sigma_1\cup\sigma_2} \to \RR^D$ such that $\phi_{\sigma_i}$ is compatible with the coloring $c_{\sigma_1} = \{l_e, \epsilon_{\sigma_i}\}$
induced on $\sigma_i$ by $c_{\sigma_1\cup\sigma_2}$. Then, if $\epsilon_{\sigma_1}=\epsilon_{\sigma_2}$,
the two points opposite to the common face in each embedded $D$-simplex do not belong to the same half-space bounded by the
hyperplane spanned by this face.

\medskip

Let us give some precisions about the notion of deficit angle:
we consider a set $\Sigma_F = \{\sigma \supset F\}$ of $D$-simplices sharing a given
$(D-2)$-face $F$ in a colored triangulation.
$\theta^{\sigma}_F$ denotes the dihedral angle of the face $F$ in the simplex $\sigma \in \Sigma_F$. The deficit angle of the face $F$ is a
function $\omega^\epsilon_F(l_e)$ of edge labels and orientations $\epsilon \equiv \{\epsilon_{\sigma}, \sigma \in \Sigma_F\}$
of the $D$-simplices surrounding $F$. It is defined by
\beq \label{defang}
\omega^{\epsilon}_F(l_e) = \sum_{\sigma \in \Sigma_F} \epsilon_{\sigma} \theta^{\sigma}_F
\eeq
Deficit angles are usually defined modulo $2\pi$. They represent the curvature localized on $(D-2)$-faces, in the sense of Regge calculus.
While each $D$-simplex can be mapped in $S^D$ (or $\RR^D$ in the flat case),
in general there is no global map, compatible with the coloring, of the complex $\Sigma_F$.
The existence of such a map requires the deficit angle (\ref{defang}) to vanish modulo $2\pi$.
We represent below, for
$D=2$, and in the case of three triangles surrounding a vertex, different configurations for which the deficit
angle at this vertex vanishes modulo $2\pi$. The complex formed by the triangles is embedded in the Euclidean plane (in the flat case).
We denote by $\epsilon$ the triplet $(\epsilon_1, \epsilon_2, \epsilon_3)$, where $\epsilon_j$ is the orientation of the triangle opposite to the point
$j$.

\begin{figure}[h]
\begin{center}
\psfrag{A}{$\epsilon = \pm (1,1,1)$}
\psfrag{B}{$\epsilon = \pm (1,-1,1)$}
\psfrag{C}{$\epsilon = \pm (1,-1,-1)$}
\includegraphics[width=11cm,height=5cm]{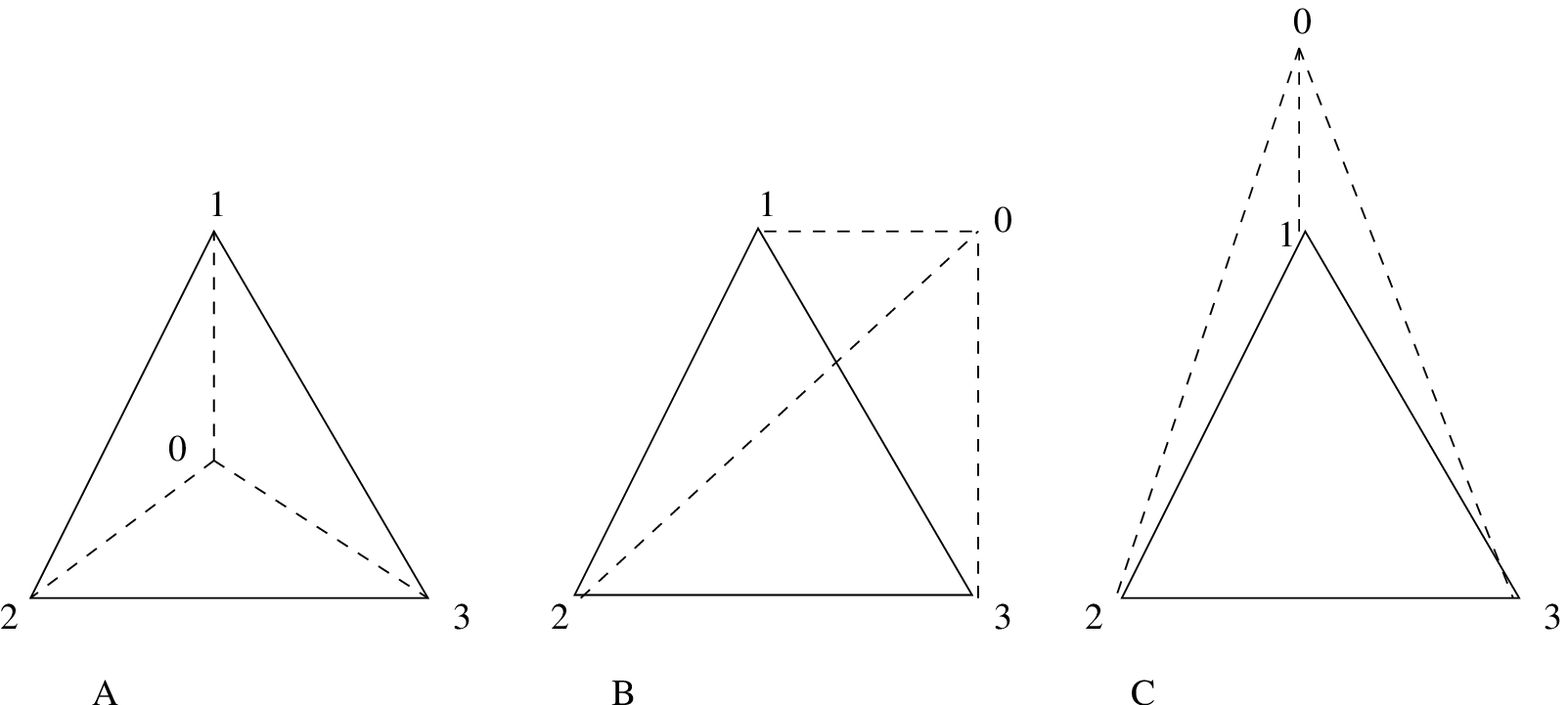}
\end{center}
\end{figure}

\medskip

We eventually define and describe some properties of the solid angles seen at vertices of a triangulation.
In a $3D$-triangulation $\Delta$, we consider a vertex $v$ and a tetrahedron $\tau$ to which $v$ belongs.
A $2d$ sphere surrounding $v$ intersects $\tau$ along a spherical triangle:
the angles of this triangle are the dihedral angles $\theta_e^{\tau}$ of the edges of $\tau$ meeting at $v$, and
its area, denoted by $A_{\tau}$,  is the solid angle seen at $v$ within $\tau$.
The spherical triangles associated to all tetrahedra sharing $v$
triangulate a surface called the \textit{link} $L_v$ of the vertex $v$. The total algebraic solid angle at $v$ is defined to be
\[ \tilde{\Omega}^{\epsilon}_v = \sum_{\tau \supset v}  A^{\epsilon_{\tau}}_{\tau} \]
where $A^{\epsilon_{\tau}}_{\tau} = A_{\tau}$ if $\epsilon_{\tau} = 1$ and $4\pi - A_{\tau}$ if $\epsilon_{\tau} = -1$. Note that
by construction $\tilde{\Omega}^{\epsilon}_v = \Omega^{\epsilon}_v \mod 4\pi$, where the RHS is defined as in section \ref{topo}.

We also define the quantity $\tilde{\omega}^{\epsilon}_e = \sum_{\tau \supset e} \tilde{\theta}^{\epsilon}_{e, \tau}$, where
$\tilde{\theta}^{\epsilon}_{e, \tau} = \theta_e^{\tau}$ if $\epsilon_{\tau} = 1$, and $2\pi - \theta_e^{\tau}$ if
$\epsilon_{\tau} = -1$. Note that this quantity is related to the deficit angle as
$\tilde{\omega}^{\epsilon}_e = \omega^{\epsilon}_e \mod 2\pi$.

Using the relation between area and angles for a spherical triangle, which reads here
$A_{\tau} = \sum_{e \supset v} \theta^{\tau}_e - \pi$,
it is straightforward to check the following equality:
\beq \label{GaussBonnet}
\frac{1}{2\pi} \left[\tilde{\Omega}_v^{\epsilon} + \sum_{e \supset v} (2\pi - \tilde{\omega}_e^{\epsilon})\right]  = \chi(L_v)
\eeq
where $\chi(L) = |\tau| - |F| + |e|$ is the Euler characteristic of the link, $|\tau|, |F|$ and $|e|$ being the number of tetrahedra,
faces and edges of $\Delta$ touching the vertex $v$, or equivalently the number of triangles, edges and vertices of the triangulation of the surface $L_v$.
If $\Delta$ triangulates a manifold, every link $L_v$ is homeomorphic to a $2$d-sphere, and therefore $\chi(L_v) = 2$. The equality (\ref{GaussBonnet})
relates the total
solid angle at $v$ with the deficit angles of the edges meeting at $v$.

Lets study the case of a vertex $0$ surrounded by four tetrahedra $1,2,3,4$. This configuration arises, for example, after a move $(1,4)$.
First, if the deficit angle $\omega^{\epsilon}_{0i}$ of the four edges $(0i)$ vanish modulo $2\pi$, then the Footnote \ref{explicit}, together with
(\ref{GaussBonnet}), yield:
\beq
\Omega^{\epsilon}_0 = 4\pi n(\epsilon), \quad \mbox{with} \quad
n(\epsilon)= \begin{cases} \epsilon_1 \quad \text{if} \quad \epsilon_i=\epsilon_j \quad \forall \, (i,j) \\
\, 0 \qquad \text{if not}
\end{cases}
\eeq
In particular, $\Omega^{\epsilon}_0 = 0 \mod 4\pi$. Now lets only suppose that the deficit angle $\omega^{\epsilon}_{04}$,
with $\epsilon = (\epsilon_1, \epsilon_2, \epsilon_3)$, of the edge $(04)$, vanishes modulo $2\pi$. With the arguments exposed
in section \ref{topo}, this means that the complex of four tetrahedra $1,2,3,4$ can be mapped in $\RR^3$, and therefore,
there exists a value of the orientation
$\epsilon^o_4$ such that the three other deficit angles vanishes modulo $2\pi$ as well, which implies
$\Omega^{\epsilon_i, \epsilon_4^o}_0 = 0 \mod 4\pi$. Now as a consequence\footnote{Except for pathological configurations whose contribution is negligible.},
one has $\Omega^{\epsilon_i, -\epsilon_4^o}_0 \not= 0 \mod 4\pi$. This
justifies the identity (\ref{gfix2}): the insertion of the function $\chi(\Omega_0^{\epsilon})$ acts as a Kronecker symbol
$\delta_{\epsilon_4, \epsilon_4^o}$, where $\epsilon_4^o = \eta\epsilon_4^{\pm}$.

\subsection{On the geometry of the simplex}

We work here in the context of spherical geometry. The results for flat space
are obtained by taking the limit where $l_{ij} \rightarrow 0$ while ratios $l_{ij}/l_{kl}$ are kept fixed.
In this limit, $G$ reduces to the square of the usual Euclidean volume, and we have the following correspondence :
\beq
\VV(e_0,\cdots, e_D) \sim D!V(l_{ij})), \quad \sin{l_{ij}}\sim l_{ij}, \quad \cos(l_{ij})-1 \sim -\frac12 l^2_{ij}.
\eeq
Abusing terminology, we will call $\VV$ the `volume' of the simplex, keeping in mind that it reduces to the volume only in the flat space limit.

\medskip

Next we will need the following result.
Let $(e_0, \ldots, e_D)$ a $D$-simplex and $L_k$ be the subspace span$\{b_0,\ldots,b_{k-1}\}$ generated
by the vectors dual to $(e_0,\ldots,e_{k-1})$. By construction $L_k$ is the subspace
orthogonal to span$\{e_{k},\cdots,e_D\}$.
We denote by $(h_0,\ldots,h_{k-1})$ the orthogonal projection of $(e_0,\cdots,e_{k-1})$ onto $L_k$.
In other words, the $h_i$ are defined to be the vectors of $\RR^{D+1}$ which satisfy the properties
\beq
h_i \cdot b_j = \delta_{ij} \quad \forall j < k, \quad \mbox{and} \quad
h_i \cdot e_j = 0 \quad \forall j \geq k
\eeq
One then has:
\begin{lemma} The following equalities hold
\beqa
\mathcal{V}(e_0,\cdots,e_D)&=&\mathcal{V}(h_0,\cdots,h_{k-1},e_{k},\cdots,e_D)\\
&=&\,\mathcal{V}(h_0,\cdots,h_{k-1})\,\mathcal{V}(e_{k},\cdots,e_D)
=\frac{\mathcal{V}(e_{k},\cdots,e_D)}{\mathcal{V}(b_0, \cdots, b_{k-1})}
\eeqa
where the function $\VV$ is defined to be $\VV(v_1,\cdots,v_n) \equiv \mbox{det}(v_1,\cdots,v_n)$ within $\mbox{span}(v_1,\cdots,v_n)$.
\end{lemma}
The first equality arises from elementary column operations on the determinant, the second one by orthogonality,
and the third one from the duality of the basis $(h_i)_{0\leq i \leq k-1}$ and $(b_i)_{0\leq i \leq k-1}$ within $L_k$.

\medskip

If one applies this result for $k =1$ one sees that $$ |b_{i}| = |h_{i}|^{-1}$$ where $|h_{i}|$ is the height of the $D-1$ face $i$, defined as
 \be \label{height}
\VV_{i}= \frac{\VV}{|h_{i}|},
\ee
$\VV$ being the $D$-simplex `volume' and $\VV_{i}$ the volume of the face $i$.
The proposition for $k=2$ leads to the identity
\be \label{sinV}
\VV \VV_{ij} = \VV_{i} \VV_{j} \sin \theta_{ij}
\ee
where $\VV_{ij}$ is the `volume' of the $D-2$ face obtained by dropping $i,j$ and $\theta_{ij}$ the dihedral angle between the faces $i$ and $j$.

Using the well known relation between the cofactor matrix and the inverse matrix one easily sees that
the dual vectors can be expressed through the derivative of  the volume:
\be
\VV b^{i}_\mu = \frac{\partial \VV}{\partial e_{i}^\mu}.
\ee
This implies the derivative formula
\be \label{derV}
\frac{\partial \VV^2}{\partial \cos l_{ij}} = -2\VV_{i}\VV_{j} \cos \theta_{ij}.
\ee
This is shown if one looks at the variation
$$
\delta \VV^2 = 2\VV^2 b^{i}_{\mu}\delta e_{i}^\mu = 2 \VV^{2}\sum_{i,j} (b^{i}\cdot b^{j}) e_{j}\cdot \delta e_{i}
=  2\VV^2  \sum_{i<j}(b^{i}\cdot b^{j} ) \delta (e_{j}\cdot e_{i})
= - 2 {\VV_{i}\VV_{j}} \sum_{i<j} \cos \theta_{ij}\, \delta \cos l_{ij}
$$
where summations over repeated indices is understood when it is not explicit, and where $i$ and $\mu$ indices are raised and lowered with flat metric.
From (\ref{sinV}, \ref{derV}), one can compute the derivative
\be\label{dertl}
\frac{\partial \theta_{ij}}{\partial l_{ij}} = \frac{\VV_{ij}\sin l_{ij}}{\VV}.
\ee
To show this, lets start from the square of  (\ref{sinV}) and derive it with respect to $\cos l_{ij}$.
Since neither $\VV_{i}, \VV_{j}$ nor $\VV_{ij}$ depends on $l_{ij}$, one obtains
$$
\frac{\partial \VV^2}{\partial \cos l_{ij}} \VV_{ij}^2 =  \VV_{i}^2 \,\VV_{j}^2\, \frac{\partial \sin^{2}\theta_{ij}}{\partial \cos l_{ij}}
= (\VV_{i}\VV_{j} \sin \theta_{ij})( 2\VV_{i}\VV_{j} \cos \theta_{ij}) \frac{\partial \theta_{ij}}{\partial \cos l_{ij}}.
$$
Using once again the identities (\ref{sinV},\ref{derV}), the RHS is equal to
$$
- \VV \VV_{ij} \frac{\partial \VV^2}{\partial \cos l_{ij}}  \, \frac{\partial \theta_{ij}}{\partial \cos l_{ij}}
=  \frac{\VV \VV_{ij}}{\sin l_{ij}} \left(\frac{\partial \VV^2}{\partial \cos l_{ij}}\right )\,\left(\frac{\partial \theta_{ij}}{\partial l_{ij}}\right),
$$
from which (\ref{dertl}) follows.
Finally we derive a last derivative identity by taking the derivative of (\ref{derV}) with respect to $\cos l_{ij}$.
With the help of (\ref{dertl}, \ref{sinV}) one obtains
\be \label{derV2}
\frac{\partial^2 \VV^2}{(\partial \cos l_{ij})^2} = -2\VV_{ij}^{2}.
\ee

\medskip

Next we will show the following

\begin{lemma} We consider a spherical $4$-simplex $(e_0,\cdots,e_4)$. We denote by $l_{ij}$ its lengths and $G \equiv G(e_0,\cdots,e_4)$ its Gram
determinant. Let $\mathcal{V}_j$ be the volume and $\epsilon_j$ the orientation of the
tetrahedron $\tau_j$ obtained by dropping the vertex $j$.
Then:
\beqa \label{3d1} \left.\frac{\pa G}{\pa l_{04}}\right|_{l_{04}^{\pm}} &=& \mp 2 \sin l_{04} \mathcal{V}_0\mathcal{V}_4 \\
\label{3d2} \left.\frac{\pa G}{\pa \omega^{\epsilon}_{04}}\right|_{l_{04}^{\pm}} &=& - 2 \epsilon_1\epsilon_2\epsilon_3
\frac{\mathcal{V}_1\mathcal{V}_2\mathcal{V}_3}{\sin l_{04}}
\eeqa
where $(l_{04}^{\pm}, \epsilon_i)$ are solution of $G = 0$ and $\omega^{\epsilon}_{04} = 0 \mod 2\pi$, with $l_{04}^- \leq l_{04}^+$.
\end{lemma}

The first equality simply arises from (\ref{derV}) and from the fact that the $4$D dihedral angle of the face $\left[123\right]$
is $\pi$ if $l_{04} = l_{04}^+$ and $0$ if $l_{04} = l_{04}^-$.

To show the second equality, we first use the Proposition $1$ for $D=3$ and $k=2$
to express the volume $\VV_i$ of the tetrahedron $\tau_i$, $i=1,2,3$, as
\beq \label{voltetra}
 \VV_i = \VV(h_j, h_k) \VV(e_0, e_4) = |h_j||h_k| \sin \theta_{ij} \sin l_{04}
\eeq
where $ijk$ is a permutation of $123$, and $|h_l|$ is the height of the point $l$ in the face $\left[04l\right]$.
Indeed, first, recall that $b_j \cdot b_k = -|b_j||b_k|\cos\theta_{jk}$, where $\theta_{jk}$ is the dihedral angle between the faces
$\left[04j\right]$ and $\left[04k\right]$ in $\tau_i$; second, writing $h_j \cdot h_k = |h_j||h_k|\cos\theta^{'}_{ij}$,
the duality of the basis $(h_j, h_k)$ and $(b_j, b_k)$ within the plane $\mbox{span}(b_j, b_k)$ implies the relation $\theta^{'}_{ij} = \theta_{ij}$.


Now, still with the Proposition $1$, the volume of the $4$-simplex factorizes as
\beq \label{varV1}
\VV = \VV(h_1, h_2, h_3) \VV(e_0, e_4) = |h_1||h_2||h_3| \VV(\tilde{h}_1, \tilde{h}_2, \tilde{h}_3) \sin l_{04}
\eeq
Since variations of the dihedral angles leave heights $|h_i|$ and length $l_{04}$ invariant, one only has to compute the variation of the quantity
$$
\VV^2(\tilde{h}_1, \tilde{h}_2, \tilde{h}_3) = \mbox{det}(\cos \theta_{ij})
$$
A first way to perform this variation is to use (\ref{derV}) with the following correspondence $l_{ij} \to \theta_{ij}$, and $\theta_{ij} \to \alpha_{ij}$,
where $\alpha_{ij}$ is the dihedral angle between edges $i$ and $j$ in the spherical triangle defined by $(\tilde{h}_1, \tilde{h}_2, \tilde{h}_3)$.
On can convince oneself that, when $G=0$, this spherical triangle is degenerate and, furthermore, $\cos\alpha_{ij} = - \epsilon_i\epsilon_j$.
Consequently, one gets
\beq \label{varV}
\delta \VV^2(\tilde{h}_1, \tilde{h}_2, \tilde{h}_3) = - 2 \epsilon_1\epsilon_2\epsilon_3 \sin\theta_{12} \sin\theta_{13} \sin\theta_{23}
(\epsilon_1\theta_{23} + \epsilon_2\theta_{13} + \epsilon_3\theta_{12})
\eeq
where the term in parenthesis is the variation of the dihedral angle $\omega^{\epsilon}_{04}$.

An other way to obtain this variation is a direct computation of the determinant. Indeed,
\beq \label{varV2}
\delta \VV^2(\tilde{h}_1, \tilde{h}_2, \tilde{h}_3) = - 2 \sin \theta_{12}(\cos \theta_{12} - \cos\theta_{13} \cos\theta_{23})\delta\theta_{12} \,
+ \mbox{perm}
\eeq
Now, since the deficit angle $\omega_{04}^{\epsilon} = \epsilon_1\theta_{23} + \epsilon_2\theta_{13} + \epsilon_3\theta_{12}$ vanishes modulo $2\pi$,
one can write
\beqa
\cos\theta_{12} - \cos\theta_{13}\cos\theta_{23} &=& \cos\epsilon_3\theta_{12} - \cos\theta_{13}\cos\theta_{23}\nonumber\\
\label{varV3}
&=&\cos(\epsilon_1\theta_{23}+\epsilon_2\theta_{13}) - \cos\theta_{13}\cos\theta_{23} = \epsilon_1\epsilon_2 \sin\theta_{23} \sin\theta_{13},
\eeqa
and insert this identity in (\ref{varV2}), to get (\ref{varV}).

Eventually, the variation of $G=\VV^2$ reads
\beq \label{G}
\delta G = - 2\epsilon_1\epsilon_2\epsilon_3 |h_1|^2|h_2|^2|h_3|^2 \sin^2 l_{04} \sin\theta_{12}\sin\theta_{13}\sin\theta_{23}\,
\delta \omega_{04}^{\epsilon}
\eeq
which, thanks to (\ref{voltetra}), can be written as
\beq
\delta G = - 2\epsilon_1\epsilon_2\epsilon_3 \frac{\mathcal{V}_1\mathcal{V}_2\mathcal{V}_3}{\sin l_{04}} \delta \omega_{04}^{\epsilon}
\eeq
This last equality finally shows (\ref{3d2}).

\section{Flatness identities for the tetrahedron} \label{6j}

We consider a flat tetrahedron $IJKL$ with edge lengths $l_{ij}$.
We define:
$\theta_I^{\,JK} \in \left[0,\pi\right]$ to be  the geometric angle   between the edges $(IJ)$ and $(IK)$ within the triangle $IJK$,
$\phi^e_{IJ}\in\left[0,\pi\right]$  to be the exterior dihedral angle of the edge $(IJ)$.
We introduce the matrices $a_I^{\,JK} \equiv a_{\theta_I^{\,JK}}$ and $h^+_{IJ} \equiv h_{\phi^e_{IJ}}$, $h^-_{IJ} \equiv h_{2\pi-\phi^e_{ij}}$, with
\[ h_{\phi} =
\left( \begin{array}{cc}
e^{\imath\phi/2}&0 \\
0&e^{-\imath\phi/2}
\end{array} \right), \quad
a_{\theta} =
\left( \begin{array}{cc}
\cos \theta/2&\sin \theta/2 \\
-\sin \theta/2&\cos \theta/2
\end{array} \right) \]
Note that, by construction, $h_{IJ}=h_{JI}$ and $a_I^{\,JK}=a_I^{\,KJ }$.
Any $SU(2)$ matrices acts as a rotation on $\mathbb{R}^3$ by adjoint action $g\cdot (x^{i}\sigma_{i})\equiv g(x^{i}\sigma_{i})g^{-1}$; as such
$h_{\xi}$ is a rotation of axis $z$ and angle $\xi$, $a_{\theta}$ a rotation of axis $y$ and angle $\theta$.

Then for all even permutation $(IJKL)$ of $(0123)$, the following vertex identities hold:
\beq \label{fid}
h^\epsilon_{IJ}  a_I^{\,LI} h^\epsilon_{IL} a_i^{\,LK}h^\epsilon_{IK}a_I^{\,KJ}= -\epsilon \mathbf{1}
\eeq
Moreover, if one considers (\ref{fid}) as an equation in $\xi_{IJ}$ where $\xi_{IJ}\in \left[0,2\pi\right]$ and $h_{IJ}=h_{\xi_{IJ}}$,
then $\xi_{IJ}=\phi^e_{IJ}$ (the exterior dihedral angle) is the only solution for $\epsilon=+1$,
and similarly  $\xi_{IJ}=2\pi-\phi^e_{IJ}$ is the only solution in $[0,2\pi]$ of (\ref{fid}) for $\epsilon=-1$.

To prove this, let us focus on the vertex $0$, and let us first give a geometrical interpretation to (\ref{fid}).
We start from a configuration of the edges meeting at $0$ such that
the edge $(01)$ is along the axis $z$, the edge  $(02) $ is in the plane $(x,z)$ and in the direction of  the axis $x$,
and the edge $(03)$ is in the direction of the
axis $y$. After the rotation $a_0^{\,21}$, the edge $(02)$ is along the axis $z$; after the rotation $h^+_{02}$ the edge $(03)$ is in the plane $(x,z)$.
The combination $h^+_{02}a_0^{\,21}$ just amounts to a cyclic permutation of the edges $(0i)$: after doing three permutations we are back where we
have started. Therefore the product (\ref{fid}) is equal to $\pm 1$.

In order to establish  our statement,
lets  compute the two sides of the equality:
\beq \label{equ}
h_{03}a_0^{32}h_{02} = -\epsilon (a_0^{13})^{-1}(h_{01})^{-1}(a_0^{12})^{-1}
\eeq
In detail the equations are
\beqa
-\epsilon e^{\imath(\xi_{01} + \xi_{02} + \xi_{03})/2} \cos\theta_0^{32}/2 &=& \cos\theta_0^{13}/2\cos\theta_0^{12}/2 -  e^{\imath\xi_{01}}\sin\theta_0^{13}/2\sin\theta_0^{12}/2 \quad (a) \nonumber \\
\epsilon e^{\imath(\xi_{01} + \xi_{02} - \xi_{03})/2} \sin\theta_0^{32}/2 &=& \sin\theta_0^{13}/2\cos\theta_0^{12}/2 + e^{-\imath\xi_{01}}\cos\theta_0^{13}/2\sin\theta_0^{12}/2 \quad (b), \nonumber
\eeqa
as well as their complex conjugate. The equations $(a)$ and $(b)$ provide four real equations whose only three are independent,
because of the unitarity condition for the matrices in (\ref{equ}). Indeed, considering $a\bar{a} + a\bar{b}$, where $\bar{a}$ denotes the complex conjugate
of the identity $(a)$, leads to the trivial equality $1=1$. Now by taking $a\bar{a} - a\bar{b}$, we get
\beq \label{car}
c_{32} = c_{12}c_{13} - \cos\xi_{01} s_{13} s_{12}
\eeq
and two similar equations obtained by permutation of indices. $c_{ij}, s_{ij}$ denote $\cos\theta_0^{\,ij}, \sin\theta_0^{\,ij}$. These equations in
$\xi_{0i}$ are those satisfied by the exterior dihedral angles. Consequently
\beq
\cos\xi_{0i} = \cos\phi^e_{0i}.
\eeq
Therefore, $\xi_{0i}= \phi^{e}_{0i}$ or $\xi_{0i}= 2\pi -\phi^{e}_{0i}$, since by hypothesis $\xi_{0i}\in [0, 2\pi]$.

In order to choose between these two possibilities we need an additional geometrical interpretation of these angles:
if we surround the vertex $0$ by a $2d$-sphere, the
three edges meeting at $0$ will intersect this sphere along three vertex of a spherical triangle, the interior of this triangle being inside the tetrahedra.
$\theta_{0}^{\,ij}$ is then the spherical length of the edge $(ij)$ of this triangle, and $\phi^{e}_{0i}$ is the exterior dihedral angle at the vertex $i$.
The area of this spherical triangle is given by
\be
A^{e}=2\pi - \phi^{e}_{01}+\phi^{e}_{02}+\phi^{e}_{03}, \quad \quad 0<A<2\pi.
\ee
We can take the imaginary part of
 $(a)$, which  implies
\beq
\epsilon  \frac{\sin \xi_{01}}{\sin A/2} = \frac{\cos \theta_0^{\,23}/2}{\sin\theta_0^{\,13}/2\sin \theta_0^{\,12}/2 } \equiv C_{1} > 0
\eeq
where $A\equiv 2\pi - \xi_{01}+\xi_{02}+\xi_{03}$.
By symmetry on can get similar identities leading to positive constants
$C_i,\, i=1\ldots3$, and therefore we find
\beq
\frac{\sin \xi_{0i}}{\sin \xi_{0j}} = \frac{C_i}{C_j} > 0 \quad \forall \, (ij)
\eeq
This last property imposes all the $\xi_{0i}$ to coincide to $ \phi^{e}_{0i}$ or all to $2\pi- \phi^{e}_{0i}$.
In the first case we have $A= A^{e}$; in the second $A= -2\pi -A^{e}$; in both cases $\sin A/2>0$.
Therefore one sees that the sign of $\sin \xi_{01} $ is $\epsilon$. This shows that the solutions  are
 $\phi_{0i}= \phi^{e}_{0i}$ when $\epsilon=+1$ and  $\phi_{0i}= 2\pi -\phi^{e}_{0i}$ when $\epsilon=-1$.

\medskip

As an application of the calculation of (\ref{equ}), let us show the expression (\ref{Jac}) for the jacobian. By considering the Euler decomposition
of the matrix $a_0^{\,12}h_{02}a_0^{\,23}\equiv h_{\phi}a_{\theta}h_{\psi}$ we write
\beq
g(\xi_{0i}) = h_{01}h_{\phi}a_{\theta}h_{\psi}h_{03}
\eeq
The Euler parameters do not depend on $\xi_{01}, \xi_{02}$. Therefore
\beqa
\extd g &=& \frac12\sin\theta \extd \theta \extd (\xi_{03}+\phi) \extd (\xi_{02} + \psi)\\
        &=& \frac12\sin\theta \extd \theta  \frac{\extd\xi_{03}}{2\pi}  \frac{\extd\xi_{02}}{2\pi}.
\eeqa
Now we know from previous calculations that the angles $\xi_{01}$ and $\theta$ are related to each other in the following way
\beq
\cos\theta = c_{12}c_{23} - \cos\xi_{01} s_{12}s_{23}
\eeq where $c_{ij}, s_{ij}$ denotes $\cos\theta_0^{ij}, \sin\theta_0^{ij}$. Consequently
\beqa
\sin\theta \extd \theta &=& 2\pi \sin\xi_{01}s_{12}s_{13}\frac{\extd \xi_{01}}{2\pi}\\
                        &=& 2\pi \frac{\VV(l_{ij})}{l_{01}l_{02}l_{03}} \frac{\extd\xi_{01}}{2\pi}
\eeqa
$\VV(l_{ij})$ is ($3!$ times) the volume of the tetrahedron whose edge lengths are $l_{ij}$. The last formula holds as long as
$\xi_{01}$ is a solution of (\ref{fid}),
since this angle is then equal, up to a sign and modulo $2\pi$, to the exterior dihedral angle of the edge $(01)$. Hence,
\beq \label{measure}
\extd g = \pi \frac{\VV(l_{ij})}{l_{01}l_{02}l_{03}} \prod_i \frac{\extd\xi_{0i}}{2\pi}
\eeq


\end{document}